\documentclass[a4paper,10pt]{article}

\usepackage{ucs}
\usepackage[utf8]{inputenc}
\usepackage{amsmath}
\usepackage{amsfonts}
\usepackage{amssymb}
\usepackage{amsthm}
\usepackage{rotating}
\usepackage{xcolor}
\usepackage[english]{babel}
\usepackage[T1]{fontenc}
\usepackage{graphicx}
\usepackage{hyperref}
\usepackage{cuted}
\usepackage{verbatim}
\usepackage{subfigure}
\usepackage{multirow}
\usepackage{bbold}
\usepackage{slashed}
\usepackage{indentfirst}
\usepackage{tcolorbox}
\usepackage[a4paper, total={7in, 10in}]{geometry}
\usepackage{bm}
\usepackage{epsfig}
\usepackage{physics}
\usepackage{graphicx}
\usepackage{caption}
\usepackage{placeins}

\newcommand{\bpsi}{\bar{\psi}}

\title{ \bf 
Magnetic field induced corrections to the NJL model coupling constant from  
vacuum polarization 
}

\author{ Thiago H. Moreira, Fabio L. Braghin 
\\
Instituto de F\'\i sica, Federal University of Goias,
Av. Esperan\c ca, s/n,
 74690-900, Goi\^ania, GO, Brazil 
}

\begin{document}

\maketitle

\begin{abstract}
Magnetic field dependent corrections for the coupling constant of the Nambu-Jona-Lasinio model 
are calculated by considering the one-loop background field method.
These coupling constants turn out to break chiral and flavor symmetries
and    they
lead to a slight improvement of the numerical  values of the up and down quark condensates
when compared to results from lattice QCD.
The corresponding magnetic field dependencies of the neutral pion and kaon masses
are also presented
 and compared with available
lattice QCD calculations.
The resulting magnetic field correction to the $\eta-\eta'$ mixing angle 
is also estimated.
\end{abstract}


PAC: 13.40.Ks, 
 12.39.Fe, 12.38.Lg, 12.39.Ki,  14.65.Bt

 \section{ Introduction}

Strong magnetic fields are expected to 
show up  in non-central heavy ions collisions  (h.i.c.)
and also in astrophysical systems such as dense stars or magnetars.
In h.i.c.   they  may reach  
$e B_0 \sim 10^{18}$ G $\sim m_{\pi}^2$ 
or  $eB_0 \sim 0.04-0.3$GeV$^2$  
from RHIC to LHC
\cite{reviews-B-qcd,parity-Bfield,hic-kharzeev,cme},
even if within a short  time interval in a limited spatial region
\cite{tuchin,skokov-Bxt,Bfield-xt-rhic}
 with recent indications that magnetic fields 
may be weaker than previously estimated although still strong
\cite{nonOhm}.
In the early Universe 
\cite{early-universe,cosmo-B}
and  in magnetars/neutron stars \cite{magnetars,eos-neutronstar,stars-B}
magnetic fields  were estimated  to have been of the order of $e B_0 \sim 10^{21}$ G and  
$10^{15}$G 
respectively.
Although
the  geometry of the magnetic fields might be time dependent and  extremely complicated,
a first theoretical analysis, by considering constant magnetic fields,  can be very useful to understand
their role in strong interactions systems - that are themselves very complicated to be treated.
Strong  magnetic fields might
lead to many different effects in different aspects of hadron dynamics,
both at the quark  and gluon level
and at the (lower energies) hadron and nuclear levels,
and  experimental evidences are currently 
 searched in different types of experiments.
The validity of the semi-classical description of magnetic field in aspects 
of h.i.c. has been tested for example in
\cite{navarra+est}.
Among these properties that might receive large contributions from the presence of the magnetic fields,
a mechanism of mass generation was predicted earlier: 
 the so called
   magnetic catalysis for which, even in the chiral limit, particles develop 
mass and that 
has been identified with the high degeneracy of the lowest Landau-level
 \cite{magn-catalysis,reviewB1,mag-catalysis-rev,mag-catalysis2}.

Usually, global properties of low energies hadrons can be suitably 
investigated by means of hadron effective models and effective field theories
whose use have been extended, more recently, for hadrons in 
strong magnetic fields
 \cite{reviewB1,reviews-B-qcd,effectmod-B,andersen-2021,md-elia,NJL-phasediagram,EPJA-2018}.
Among the successful  QCD  effective models,  the Nambu-Jona-Lasinio (NJL) model is known to
reproduce, and eventually to predict, many 
observables for the hadron structure and dynamics under different conditions 
\cite{NJL,NJL1,NJL2,kunihiro-hatsuda,andersen-2021}.
Several approaches have been already employed to describe how
the NJL model coupling constant might be obtained in terms of  QCD degrees of freedom 
in the vacuum
\cite{kleinert,chilenos,kondo,coimbra-etc,weise-etal,PRD-2014}
or to understand further how those degrees of freedom contribute for the NJL-model parameters
\cite{pcosta-etal,PRD-2021,JPG-2022}.
Lately, lattice QCD  provided results for hadron observables in a finite strong magnetic field
 were also 
used to test the predictions of NJL model  in such conditions and,
eventually, this type of comparison may  favor 
 an improvement of its
the predictive power 
\cite{bali-etal-2018,endrodi+marko,NJL-meet-latt}.
With such comparisons, one might also obtain knowledge on 
how the parameters of the effective models might be related to more fundamental degrees of 
freedom from QCD.
It has been   envisaged  that 
the NJL coupling constant
 might receive magnetic field contributions, $G(B)$  \cite{gastao-etal-PRC}, 
because of quark and gluon interactions \cite{asymfree-igor-etal}.
This effect  
contributes for the improvement of the description of the quark-antiquark chiral condensate
as function of the magnetic field
\cite{NJL-meet-latt}.
Among  important hadron observables, 
the   light pseudoscalar mesons have a special role in 
the Strong Interactions  since they are  the quasi-Goldstone bosons of 
the Dynamical Chiral Symmetry Breaking (DChSB).
Their masses  in the vacuum are associated basically to the explicit 
breaking of chiral symmetry and its amplification due to the DChSB.
Their behavior under strong magnetic fields 
was investigated  in the last years.
Several 
 results were obtained from 
calculations with the NJL-model in strong magnetic fields by assuming  
 $G(B)$ or not
 \cite{gastao-etal-PRC,xu-etal,hattori-etal,avancini-etal,avancini-etal2,mao-etal,coppo-etal-2,gauge-ind,chao-etal,nonet-NJL-B,g-cao}.
Lattice QCD has provided  few estimations  for the behavior of hadron properties under 
strong magnetic fields, being that 
a small difference was found when comparing earlier different lattice fermions
\cite{bali-etal-2018,endrodi+marko,htding-etal-latt}.
Although this dependence of $G(B)$
 has been attributed to the quark-gluon running coupling constant 
dependence on the magnetic field \cite{asymfree-igor-etal},
  we believe that 
 other mechanisms can contribute.
Besides that, one might be interested in understanding more precisely the 
role of  {\it needed}  degrees of freedom of the more fundamental theory
for defining the NJL-model parameters by 
articulating further the model itself and their parameters.

In the present work we employ the background field method (BFM) \cite{BFM,EPJA-2018}
to compute the contribution of the  quark-polarization under strong magnetic field
for the NJL-coupling constant.
This work extends the more restricted calculation for weak magnetic field and $SU(N_f=2)$
presented in 
\cite{PRD-2016} and, besides that, estimations for its effects on the neutral pion and kaon masses
are presented.
The dependencies of the u, d and s quark condensates and of 
the $\eta-\eta'$ mixing angle 
\cite{mixing1,PDG,PRD-2021,JPG-2022}
on the magnetic field are also calculated.
The  auxiliary field method (AFM) will be considered, as usually, being that the  scalar field
allows for the DChSB although a chiral rotation is performed to eliminate 
the corresponding meson degree of freedom which seems  absent in the light
 hadron spectrum.
For zero magnetic field it has been shown that the choice of the regularization method 
is little important for the light hadron observables \cite{kohyama-etal}
and an investigation for the role of different regularization schemes under finite B 
has also been carried out \cite{debora-etal,sidney-reg-presc}.
A magnetic field independent regularization is chosen
 for the  covariant four dimensional cutoff regularization.
We make use of the (more convenient) 
proper-time representation for the magnetic field contribution 
for the quark propagator that is 
ultraviolet (UV) finite.
The B-dependence of the results are  guided strictly by the behavior of the 
effective masses from the gap equations, and eventually B-dependent coupling constants.
However, since we are concerned with the relative role of the magnetic field dependent coupling constants
with respect to the original NJL-coupling constant, $G_0$, - i.e. to analyze 
the relative influence due to the magnetic field -  the role of the choice of the 
regularization scheme (in particular for the vacuum part of the 
equations) may be expected to  be 
relatively small.
By resolving the coefficients of a large quark 
mass expansion of the quark determinant
in the background quark currents,
for a zero order derivative expansion,
the magnetic-field dependent
corrections for the NJL-coupling constant, $G_{ij}(B)$, are obtained mostly analytically.
These coupling constants turn out to be strongly flavor dependent.
The fitting of the parameters of the parameters of the resulting model,
with $G_{ij}$, is  done  by means of the usual
observables in the vacuum, neutral mesons masses $M_{\pi^0}, M_{K^0}$ 
and the
 decay constants  $F_{\pi}$ and $F_{K}$.
The work is organized as follows.
In the next section the sea-quark determinant 
is presented
in the presence of background
 scalar and pseudoscalar quark currents and local pseudoscalar and scalar
auxiliary fields. 
In Section (\ref{sec:correctedNJL})
the corrected NJL-model, wtih  $G_{ij}(B)$,
 is considered for the calculation of
the neutral pion and kaon masses as functions of the magnetic field.
In section (\ref{sec:numerics}) numerical results are presented
for the quark effective masses, scalar and pseudoscalar magnetic field dependent 
(corrected) coupling constants, quark-antiquark chiral condensates and
neutral pion and kaon masses.
Besides that, a magnetic field correction to the
 $\eta-\eta'$ mixing angle 
 will be also calculated for 
different behaviors of the  magnetic field dependencies of 
the $\eta-\eta'$ mass difference that is, currently, also unknown.
Finally in section (\ref{sec:summary}) a Summary with a discussion is presented.

\section{
Background field method, sea quark determinant and gap equation
}

 The following generating functional will be considered:
$$
Z [ J, \bar{J}] = N  \int {\cal D}[\bpsi,\psi]
e^{ i \int_x ( {\cal L} +\bpsi J + \bar{J} \psi) },
$$
where the NJL-model  Lagrangian density for the 
 minimal  coupling for a background electromagnetic field
 can be  written as:
 \begin{eqnarray} \label{NJL-standard}  
{\cal L}  = 
\bar{\psi} \left( i \gamma \cdot { D}
- m_f \right) \psi 
+
\frac{G_0}{2}\left[
(\bpsi  \lambda^i \psi)^2
+ (\bpsi i \gamma_5 \lambda^i \psi)^2
\right]
,
\end{eqnarray}
where 
 $i,j,k=0,...(N_f^2-1)$  stand 
for flavor indices in the adjoint representation, $m_f$ stand for 
the current quark mass matrix element wherein
$f=u,d$ and $s$ for the fundamental
representation
and  
 the sums in color, flavor and Dirac indices are implicit.
The covariant quark derivative  is:
$D = D_{\mu} = \partial_\mu \delta_{ij} - i e Q_{ij} A_{\mu}$
for the diagonal matrix $\hat{Q}
=  diag(2/3, -1/3, -1/3)$.

 Next we apply the one loop
  Background Field Method (BFM)
\cite{BFM,EPJA-2018}
according to which
bilinears of the quark field,  $\bpsi \Gamma \psi$
where $\Gamma$ stands for Dirac, color  or flavor operators,
  are split into
(constituent quark)  background field ($\psi_1$) that will become 
{\it quasi-particles} of the model 
 and the  quantum quark field ($\psi_2$) that will form 
mesons and the chiral condensates and which  will be integrated out.
It can be written
\begin{eqnarray} \label{split-Q} 
\bpsi \Gamma^q \psi &\to& (\bpsi \Gamma^q \psi)_2 + (\bpsi \Gamma^q \psi)_1.
\end{eqnarray}
This separation 
   preserves chiral symmetry
and it may  not correspond to  a simple mode separation  of low and high energies 
which might be a very restrictive assumption
and what would 
involve an energy  separation  scale.
Whereas the overall  method employed is inspired in the 
usual constant background field method, one step further can be given with the derivative expansion
that allows to compute a whole effective action  \cite{EPJA-2016,PRD-2019,JPG-2020b}.

The interaction term of the NJL model, $\Omega$,  split 
in separated terms for $\psi_1$ and $\psi_2$ ($\Omega_1, \Omega_2$)
and those with both bilinear  of $\psi_1$ and $\psi_2$.
The interaction $\Omega_2$
will be treated by
making use of the 
auxiliary field method (AFM)  by introducing 
a set of auxiliar scalar and pseudoscalar fields,
$S_i\equiv S_i (x)$ and $P_i\equiv P_i(x)$ for $i=0,1,..N_f^2-1$,
 \cite{kleinert}.
These auxiliary fields  might be introduced by multiplying the generating functional 
by the following normalized Gaussian integrals:
\begin{eqnarray}
 1 &=& N \int D[S] D[P_i]
 e^{- \frac{i}{2 G_0}  
\int_x      \left[ (S_i  - G_0  j^S_{i,(2)})^2 +
(P_i -  G_0  j^{P}_{i,(2)} )^2 \right]}
,
\end{eqnarray}
where
$\int_x = \int d^4x$ and  the scalar and pseudoscalar currents were defined as:
$j_S^{i,(2)} = \bar{\psi} \lambda_i \psi $ and $j_P^{i,(2)} = \bar{\psi} \lambda_i  i \gamma_5 \psi$.

With these auxiliary fields, the quark field $\psi_2, \bar{\psi}_2$ can be quantized,
and an
effective action  for background quarks   and 
auxiliary fields canonically normalized, 
 is obtained.
From here on, we can omit the index for quark background field.
By considering the identity
$\det A = \exp \; Tr \; \ln (A)$, 
the resulting model can be written as:
\begin{eqnarray} \label{Seff-det}  
S_{eff}   &=&   -  i  \; Tr  \; \ln \; \left\{
- i \left[ {S_0^B}^{-1}  
+ \Xi  
+
   G_0
\lambda^i \left[  (\bpsi \lambda_i  \psi) 
+ i  \gamma_5 (\bpsi i \gamma_5  \lambda_i \psi) \right] \right]
 \right\} 
\nonumber
\\
&+& \int_{x} \left\{ 
 \bar{\psi} \left( i \gamma_{\mu} { D}^{\mu} 
- m \right)\psi
+
\frac{G_0}{2}
\left[
(\bpsi  \lambda^i \psi)^2
+ (\bpsi i \gamma_5 \lambda^i \psi)^2
\right]
-  \frac{1}{2 G_0}   \left[ S_i^2 +
P_i^2 \right]
 \right\} ,
\end{eqnarray}
where 
$Tr$ stands for traces of discrete internal indices 
and integration of  space-time coordinates and 
the following quantities have been defined:
\begin{eqnarray}
{S_0^B}^{-1}  &=& \left(  i \slashed{D} -  m_f
\right) ,
\\
\Xi  &=& \left(  S \cdot \lambda + i P \cdot \lambda \right)
,
\end{eqnarray}
where $S_0^B$ is 
the   free quark propagator with its coupling to the electromagnetic field. with
 $\slashed{D} =  \gamma^\mu \cdot D^\mu$.
Therefore $\Xi$ 
provides 
 the auxiliary fields  coupling to quarks.

Since the auxiliary fields are unknown, 
an extremization of eq.  (\ref{Seff-det}) yields the usual gap equations
and provide a determination of the auxiliary fields at a mean field level,
$\bar{S}_i$.
The  solutions of the scalar fields
for the gap equations
have been investigated in many works both in the vacuum   and 
under constant weak and strong magnetic fields, 
to quote few works:  \cite{Meff-B,chiral-cond,cond-B,reviewB1}.
The magnetic field is known to increase the effective mass, even if the current Lagrangian quark mass 
is zero,  which is known as the 
magnetic catalysis effect \cite{magn-catalysis,reviewB1,mag-catalysis-rev,mag-catalysis2}.
The resulting gap equations  for the set of scalar auxiliary fields
corresponding to the diagonal flavor generators, $S_0, S_3$ and $S_8$,
 can be written as:
\begin{eqnarray}  \label{gap1}
S_i \equiv \bar{S}_i &=& - i \;  G_0 \; Tr \; \lambda_i \;  S^{(B)},
\end{eqnarray}
where $S^{(B)}$ (defined below) 
takes into account possible non zero expected value in the vacuum for the auxiliary fields.
The  corresponding equations for the pseudoscalar fields, at zero magnetic field, 
  must be a trivial one to 
enforce the scalar nature of the vacuum. 
The scalar auxiliary field mean field makes possible the generation of  (effective) mass for 
the constituent quarks, such that in the fundamental representation one 
has $M^*_f = m_f + \bar{S}_f$.
The quark propagator in a background magnetic field was calculated by considering the 
Schwinger proper time 
method and it is shown explicitly in Appendix (\ref{sec:q-prop})
In the absence of (background) quark currents and  auxiliary fields for mesons
 the celebrated Euler Heisenberg effective action  can be recovered
from Eq. (\ref{Seff-det})
\cite{cond-B,EH1936,schwinger51,IZ}.

 \subsection{ GAP equation in magnetic field}

The non trivial solution for the scalar variables 
 lead to diagonal contributions for the
fundamental representation, i.e. $\bar{S}_u, \bar{S}_d$ and $\bar{S}_s$.
From here on  the quark masses become effective masses such 
that one can write:
\begin{eqnarray}
{S^B}^{-1}_f  = \left(  i \slashed{D}_f -  M^*_f
\right),
\end{eqnarray}
where $ M^*_f = m_f + \bar{S}_f$ and 
the different minimal photon couplings to u, d and s quarks were written above in $\slashed{D}_f$.
The gap equation for the effective quark masses can be written as
\begin{equation}
    M_f^*=m_f- 2G_0\expval{\bar{\psi}_f\psi_f},
\end{equation}
where $\expval{\bar{\psi}_f\psi_f}=-i\textrm{tr}_{DC}S_f^{B}(0)$ is the chiral condensate in the mean field approximation. 
Here $\textrm{tr}_{DC}$ denotes the trace over Dirac and color indices, and $S_f^{B}(x-y)$ stands for the quark propagator of flavor $f$ in the presence of a uniform magnetic field. 
A magnetic field independent regularization will be adopted \cite{debora-etal,sidney-reg-presc}.
The UV divergent part can be separated to correspond to the vacuum contribution whereas
the explicitly  magnetic field dependent part is UV finite. 
The regulariation scheme considered for the UV divergent part will be the
 four-momentum cutoff ($\Lambda$) in Euclidean space.
By using the proper time representation for the magnetic field dependent part  we find
\begin{equation} \label{Gap-equation-magnetic}
    \begin{split}
        M_f^*&=m_f+\frac{G_0N_cM_f^*}{2\pi^2}\qty[\Lambda^2-{M_f^*}^2\ln\qty(\frac{\Lambda^2+{M_f^*}^2}{{M_f^*}^2})] 
\\
        &+\frac{G_0N_c{M_f^*}}{2\pi^2}\left[ {M_f^*}^2\qty(1-\ln\frac{{M_f^*}^2}{2\abs{q_fB}})+\abs{q_fB}\ln\frac{{M_f^*}^2}{4\pi\abs{q_fB}}+2\abs{q_fB}\ln\Gamma\qty(\frac{{M_f^*}^2}{2\abs{q_fB}})\right]
\\
&= M_{f,0}^{*} 
+ \frac{G_0N_c{M_f^*}}{2\pi^2}\left[ {M_f^*}^2\qty(1-\ln\frac{{M_f^*}^2}{2\abs{q_fB}})+\abs{q_fB}\ln\frac{{M_f^*}^2}{4\pi\abs{q_fB}}+2\abs{q_fB}\ln\Gamma\qty(\frac{{M_f^*}^2}{2\abs{q_fB}})\right].
    \end{split}
\end{equation} 
We remark that the divergences were isolated
 into the vacuum term 
 before introducing the regularization parameter. 
Although our departure point was the proper time representation for the propagator, isolating the divergences into the vacuum contribution allowed us to use other regularization scheme than the regularization in proper time since the pure magnetic contribution introduces no new divergences.

\subsection{ Magnetic field-dependent corrections to the coupling constant }
  
Since we are interested in the dynamics of quarks by means of their currents from here on
the auxiliary fields will be neglected.
By expanding the quark determinant in a large quark effective  mass expansion
  in terms of the quark field bilinears, in a zero order 
derivative expansion, we find the first order term to be given by
\begin{equation}
    S_{\textrm{det}}^{(1)} = - 2G_0\sum_{f=u,d,s}\int_x\textrm{tr}_{DC}\qty[iS_f^{B}(p)]\bar{\psi}\psi,
\end{equation}
with $S_f^B(p)$ representing the quark propagator in momentum space in the presence
 of the uniform magnetic field   $B$,
that is exhibited in Appendix (\ref{sec:q-prop}).
These terms 
produce a correction to the quark masses 
that is the same as  the gap equation, Eq. \eqref{Gap-equation-magnetic}.

The second order terms of the large quark mass expansion provides fourth order quark interactions.
After resolving coupling constants in the very long-wavelength limit for
 the zero order derivative expansion,  results
are the following:
\begin{eqnarray}
{\cal L}_{1loop} &=&
\frac{\bar{G}_s^{ij}(B)}{2} (\bar{\psi} \lambda_i \psi)  (\bar{\psi} \lambda_j \psi) + 
\frac{ \bar{G}_{ps}^{ij}(B)}{2} (\bar{\psi} i\gamma_5 \lambda_i \psi)  
(\bar{\psi} i\gamma_5 \lambda_j \psi),
\end{eqnarray}
where:
\begin{eqnarray} \label{Gs-B}
    \bar{G}_s^{ij}(B) &\equiv&  G_0^2   \Pi^{s}_{ij} (B)
= 
i
G_0^2\int\frac{d^4p}{(2\pi)^4}\textrm{tr}\qty[S_f^{B}(p)\lambda^iS_g^{B}(p)\lambda^j],
\\ \label{Gps-B}
    \bar{G}_{ps}^{ij}(B)
&\equiv&     G_0^2  \Pi^{ps}_{ij} (B)
= i
G_0^2\int\frac{d^4p}{(2\pi)^4}\textrm{tr}\qty[S_f^{B}(p)\lambda^i
i\gamma_5 S_g^{B}(p)\lambda^j i\gamma_5].
\end{eqnarray}
All these
coupling constants are written as combinations of integrals of each quark propagator 
for which a change of representation for the coupling constants is presented 
in the 
Appendix (\ref{sec:GijGfg}).
These coupling constants obviously  break chiral and flavor symmetries
and they have the same dimension of the NJL-coupling constant, GeV$^{-2}$.
 By using the proper time representation for the quark propagator in momentum space and considering only the polarization functions that involve quark flavors with the same electric charge, being that in those cases the Schwinger phases cancel out,  it is possible to separate the contributions from the vacuum (zero magnetic field)
and the B-dependent contributions similarly to the quark propagator.
By separating each of the couplings for given $i,j$ in terms of the related contributions from
internal quark propagators $f,g$
it is obtained for each component with equal electric charges $q_f=q_g$:
\begin{equation}
    \begin{split}
        \Pi_{fg}^{\substack{\textrm{s}\\\textrm{ps}}}(B)&=
\Pi_{fg}^{\substack{\textrm{s}\\\textrm{ps}}}(B=0)+
\tilde{\Pi}_{fg}^{\substack{\textrm{s}\\\textrm{ps}}} (B)
= \Pi_{fg}^{\substack{\textrm{s}\\\textrm{ps}}}(B=0)
+
\frac{N_c\abs{q_fB}}{2\pi^2}\int_0^\infty\int_0^\infty dsdr\,\frac{e^{-s{M_f^*}^2-r{M_g^*}^2}}{s+r} \\
        &\times\left[ \frac{1\mp {M_f^*}{M_g^*}(s+r)}{(s+r)\tanh\qty(\abs{q_fB}(s+r))} +\frac{\abs{q_fB}}{\sinh^2\qty(\abs{q_fB}(s+r))}-\frac{2\mp {M_f^*}{M_g^*}(s+r)}{\abs{q_fB}(s+r)^2}\right] .
    \end{split}
\end{equation}
The vacuum contributions for the flavor symmetric model were analyzed in \cite{PRD-2014,PLB-2016}
and for non degenerate quark masses in \cite{PRD-2021,JPG-2022}.

By making the change of variables
\begin{equation}
    s=\frac{u}{2}\qty(1+v),\,\,\,\,\,\,\,\,\,\,r=\frac{u}{2}\qty(1-v),
\end{equation}
with $0\leq u<\infty$ and $-1\leq v\leq1$, so that $dsdr=(u/2)dudv$, we obtain:
\begin{equation}
    \tilde{\Pi}_{fg}^{\substack{\textrm{s}\\\textrm{ps}}}(B)
=\frac{N_c\abs{q_fB}}{2\pi^2}\int_0^\infty du\int_{-1}^1dv\,\frac{e^{-\frac{u}{2}\qty(1+v){M_f^*}^2-\frac{u}{2}\qty(1-v){M_g^*}^2}}{2} \left[ \frac{1\mp u{M_f^*}{M_g^*}}{u\tanh\qty(\abs{q_fB}u)} +\frac{\abs{q_fB}}{\sinh^2\qty(\abs{q_fB}u)}-\frac{2\mp u{M_f^*}{M_g^*}}{\abs{q_fB}u^2}\right] 
\end{equation}
for the pure magnetic contribution to the polarization functions.
 The proper time integrals can be computed in closed form for
the diagonal couplings $f=g$, yielding
\begin{equation*}
    \begin{split}
        \tilde\Pi_{ff}^{\substack{\textrm{s}\\\textrm{ps}}}(B)=\frac{N_c{M_f^*}^2}{2\pi^2}&\left[ 1+\frac{\abs{q_fB}}{{M_f^*}^2}\ln\qty(\frac{{M_f^*}^2}{4\pi\abs{q_fB}})+\frac{2\abs{q_fB}}{{M_f^*}^2}\ln\Gamma\qty(\frac{{M_f^*}^2}{2\abs{q_fB}})\right. \\
        &\left. +\qty(1\pm1)\psi\qty(\frac{{M_f^*}^2}{2\abs{q_fB}})-\qty(2\pm1)\ln\qty(\frac{{M_f^*}^2}{2\abs{q_fB}})+\qty(1\pm1)\frac{\abs{q_fB}}{{M_f^*}^2}\right] .
    \end{split}
\end{equation*}

Therefore, we have
\begin{equation} \label{Gs-B}
    \begin{split}
        \bar{G}_{ff}^{\,\rm s}(B)&= 
\frac{G_0^2N_c{M_f^*}^2}{2\pi^2}\left[ 1+\frac{\abs{q_fB}}{{M_f^*}^2}\ln\qty(\frac{{M_f^*}^2}{4\pi\abs{q_fB}})+\frac{2\abs{q_fB}}{{M_f^*}^2}\ln\Gamma\qty(\frac{{M_f^*}^2}{2\abs{q_fB}})\right. \\
        &\,\,\,\,\,\,\,\,\,\,\,\,\,\,\,\left. +2\psi\qty(\frac{{M_f^*}^2}{2\abs{q_fB}})-3\ln\qty(\frac{{M_f^*}^2}{2\abs{q_fB}})+2\frac{\abs{q_fB}}{{M_f^*}^2}\right] ,
    \end{split}
\end{equation}
\begin{equation}  \label{Gps-B}
    \bar{G}_{ff}^{\,\rm ps}(B)= 
\frac{G_0^2N_c{M_f^*}^2}{2\pi^2}\qty[ 1+\frac{\abs{q_fB}}{{M_f^*}^2}\ln\qty(\frac{{M_f^*}^2}{4\pi\abs{q_fB}})+\frac{2\abs{q_fB}}{{M_f^*}^2}\ln\Gamma\qty(\frac{{M_f^*}^2}{2\abs{q_fB}})-\ln\qty(\frac{{M_f^*}^2}{2\abs{q_fB}})],
\end{equation}
where $\Gamma(x)$ is the Gamma function and $\psi(x)$ is the Euler psi function.

Another case of interest is the one of the couplings $G_{ds}^{\,\rm s}(B)$ and $G_{ds}^{\,\rm ps}(B)$, which involve quarks of different flavors but with same electric charge. We have
\begin{equation} \label{Gsds-Gpsds}
    \begin{split}
        \bar{G}_{ds}^{\substack{\textrm{s}\\\textrm{ps}}}(B)&= 
\frac{G_0^2N_c\abs{q_dB}}{2\pi^2}\int_0^\infty du\int_{-1}^1dv\,\frac{e^{-\frac{u}{2}\qty(1+v){M_d^*}^2-\frac{u}{2}\qty(1-v){M_s^*}^2}}{2} \\
        &\,\,\,\,\,\,\,\,\,\,\,\,\times\left[ \frac{1\mp u{M_d^*}{M_s^*}}{u\tanh\qty(\abs{q_dB}u)} +\frac{\abs{q_dB}}{\sinh^2\qty(\abs{q_dB}u)}-\frac{2\mp u{M_d^*}{M_s^*}}{\abs{q_dB}u^2}\right] ,
    \end{split}
\end{equation}
where now the proper time integrals need to be solved by using numerical methods. 
Note that, the divergence of the integrals above,  parameterized in the UV cutoff $\Lambda$,
appears only in the vacuum contributions $G(B=0)$ or $\Pi(B=0)$.
The
the  difference between the scalar and pseudoscalar couplings 
is directly a consequence of  chiral symmetry breaking effect in the coupling constants
at the one loop level.
For the cases addressed above it follows:
\begin{eqnarray} \label{Gs-Gps} 
\bar{G}^{fg}_{csb} (B) &\equiv&
\bar{G}_{ps}^{fg} (B) - \bar{G}_s^{fg} (B) 
\nonumber
\\
 &=&
2 G_0^2 
\frac{N_c\abs{q_fB}}{2\pi^2}\int_0^\infty\int_0^\infty dsdr\,\frac{e^{-s{M_f^*}^2-r{M_g^*}^2}}{s+r} 
      \left[ \frac{ {M_f^*}{M_g^*}(s+r)}{(s+r)\tanh\qty(\abs{q_fB}(s+r))}
-\frac{ {M_f^*}{M_g^*}(s+r)}{\abs{q_fB}(s+r)^2}\right] .
\end{eqnarray}
Although the magnetic field dependence of this quantity is not necessarily small,
it will be neglected in most of calculations as explained  below.
The overall behavior of the pseudoscalar coupling constants is 
very different from the needed behavior that describes results 
from lattice QCD.

\subsection{ Fitting of the model parameters at $B=0$ and contributions from $B\neq 0$}

 The  parameters of the model are $G_0, m_u, m_d$ and $m_s$ with the additional
need of fixing the UV cutoff $\Lambda$.
By adopting a coupling constant $G_0 = 9.76$ GeV$^{-2}$ 
the following masses and weak 
decay constants were considered in the vacuum to fix these parameters 
$M_{\pi^0}, M_{K^0}, F_\pi$ and $F_K$, that are written below.

The  pseudoscalar mesons masses ($M_{ps}$) in the framework of the standard NJL model 
 (\ref{NJL-standard})
are obtained from the  Bethe Salpeter equation at the Born approximation
by means of the following equations:
\begin{eqnarray} \label{BSE-G0}
\left. 1 - G_0 \Pi_{ps}^{ij} (P^2)  \right|_{P^2 =  M_{ps}^2 } = 0,
\end{eqnarray}
where neutral pion and kaon are  obtained  respectively  with $ij=33$  and $ij=66,77$.
The polarization tensor was rotated back to the Minkowski space
and it needs to be computed for on shell meson, in the limit of  zero three-momentum.
The  corresponding integrals are given by:
\begin{eqnarray} 
\Pi_{\textrm{ps}}^{33}(P^2) &=&  \frac{1}{2}\left[ \Pi_{\textrm{ps}}^{uu}(P^2) + \Pi_{\textrm{ps}}^{dd}(P^2)\right] ,
\\
\Pi_{\textrm{ps}}^{66}(P^2)  &=&  \Pi_{\textrm{ps}}^{ds}(P^2) .
\end{eqnarray}
These pseudoscalar polarization tensors  as functions of the 
energy of the meson
 can be written as:
\begin{equation} \label{piff-pifg}
    \begin{split}
        \Pi_{\textrm{ps}}^{fg}(P^2)&=
\frac{1}{2} \left[ \frac{ M_{f,0}^* - m_f}{M_f^*} + 
\frac{ M_{g,0}^* - m_g}{M_g^*} \right]
\\
        &+\frac{N_c}{4\pi^2}\qty[P^2-\qty({M_f^*}-{M_g^*})^2]\int_0^1dx\,
\qty{\ln\qty[\frac{\Lambda^2+D_{fg}^2(P^2)}{D_{fg}^2(P^2)}]+\frac{D_{fg}^2(P^2)}{\Lambda^2+D_{fg}^2(P^2)}-1} \\
        &+\frac{N_c\abs{q_fB}}{2\pi^2}\int_0^\infty\int_0^\infty dsdr\,\frac{e^{-s{M_f^*}^2-r{M_g^*}^2+\frac{sr}{s+r}P^2}}{s+r} \\
        &\times\qty[\frac{1+{M_f^*}{M_g^*}(s+r)+\frac{sr}{s+r}P^2}{(s+r)\tanh\qty(\abs{q_fB}(s+r))}+\frac{\abs{q_fB}}{\sinh^2\qty(\abs{q_fB}(s+r))}-\frac{2+{M_f^*}{M_g^*}(s+r)+\frac{sr}{s+r}P^2}{(s+r)^3}],
    \end{split}
\end{equation}
where $M^*_{f,0}$ was defined in Eq. (\ref{Gap-equation-magnetic}) and
 $D_{fg}^2(P^2)=-x\qty(1-x)P^2+x{M_f^*}^2+\qty(1-x){M_g^*}^2$.
Note that, again, the UV divergent part is written separately, as  the vacuum term,
 from the magnetic field contributions.

The   charged pion and kaon   weak decay
 constants, for a meson structure of quark antiquark $f, g$,
are  given by 
 \cite{NJL1,NJL2}:
\begin{eqnarray}
F_{ps} =  \frac{N_c  \; G_{qq PS}}{4}\; \int \frac{ d^4 q}{(2 \pi)^4}
 Tr_{F,D} \left[ \gamma_\mu \gamma_5 
\lambda_i \; S_{f} (q + P/2) \lambda_j S_{g} (q - P/2) \right],
\end{eqnarray}
where 
$i,j$ are the associated flavor indices as discussed for eq. (\ref{BSE-G0}).
The meson-quark coupling can be obtained as the residue of the pole of the 
BSE will be calculated in the limit of zero four momentum as:
\begin{eqnarray}
G_{qq PS} =  \left( \frac{ \partial \Pi_{ij} (P^2)
 }{\partial P_0^2 } \right)^{-2}_{(P_0, \vec{P})\equiv 0},
\end{eqnarray}
where the flavor indices are tied with the quantum numbers of the 
meson $PS$, $\pi^+$ with $i,j=(1,1),(2,2)$ and $K^+$ with $i,j=(4,4),(5,5)$.

\section{ Corrected NJL-model}
\label{sec:correctedNJL}

Now consider the NJL corrected with magnetic field dependent coupling constants
found above
as given by:
 \begin{eqnarray} \label{NJL-Gij}  
{\cal L}  &=& 
\bar{\psi} \left( i \gamma \cdot { D}
- m \right) \psi 
+
\frac{G_0 \delta_{ij} + \bar{G}_s^{ij} (B) }{2}
\left[
(\bpsi  \lambda^i \psi) (\bpsi  \lambda^j \psi)  
+ ( \bpsi i \gamma_5 \lambda^i \psi) ( \bpsi i \gamma_5 \lambda^j \psi) \right] + {\cal O}_{chsb}
\end{eqnarray}
 where  $\bar{G}_{ij}^s (B)$ and 
$\bar{G}^{ij}_{csb} (B)$ were written in 
Eqs.  (\ref{Gs-B}) and  (\ref{Gps-B}) and ${\cal O}_{chsb}$ 
are the chiral symmetry breaking corrections for the 
pseudoscalar coupling constants discussed above and neglected from here on.
However, the zero magnetic field limit of $G_{ij}^{s,ps}(B=0)$ contributes for $G_0$
making the overall normalization of the coupling constant ambiguous
and this would have consequences for the calculation of observables 
for which we are interested, however,
 in investigating only the magnetic field dependence.
Therefore the B-dependent coupling constant considered in the Lagrangian above
will be exclusively the B-dependent part of Eq. (\ref{Gs-B}).
 In principle the scalar interactions contribute for the gap equations and the 
pseudoscalar couplings $G_{ij}^{ps}$ can be expected to be those by which 
  the bound state pseudoscalar mesons are formed according to the BSE.
Because of the completely different behavior of the scalar and pseudoscalar couplings
and of the fact that the scalar coupling constant helps to improve the 
behavior of the quark-antiquark condensates with the B-field that is shown below,
we decided to analyze rather the effects of the scalar coupling by neglecting
 the difference between them.
So we assume that, for some unknown reason maybe related
to the level of approximation in which the one loop quark determinant and its expansion 
rely,
 the different between the scalar and pseudoscalar
couplings $G_{ij}^s$ and $G_{ij}^{ps}$ should be considerably smaller
and the pseudoscalar coupling constants would have its behavior changed considerably to 
be similar to the scalar ones, 
i.e. $\bar{G}_{ij}^s (B) \sim \bar{G}_{ij}^{ps} (B)$ given by Eq.  (\ref{Gs-B}).
Next, the usual logics applied to the NJL model 
must be used again and so the gap equations must be recalculated.
This procedure can be done repeatedly until the resulting effective masses, $M_f^*$,
and coupling constants, $G_{ij}^{s,ps}$, converge.

The auxiliary field method for the corrected model will be presented by neglecting all the 
mixing-type  interactions $G_{i \neq j}$   
that are considerably smaller than the diagonal ones. 
The corresponding unit integral of the auxiliary fields can be written as:
\begin{eqnarray}
 1 &=& N \int D[S_i] D[P_i]
 e^{- \frac{i}{2 G_{ij}}  
\int_x      \left[ (S_i  - G_{ik}  j^S_{k} )   (S_j - G_{jm}  j^S_{j} ) +
(P_i -  G_{ik}  j^{P}_{k} ) (P_j -  G_{jm}  j^{P}_{m} ) 
\right]}
,
\end{eqnarray}
where
$G_{ij} = G_{ij}^s = G_0 \delta_{ij} + \bar{G}^{s}_{ij} (B)$.
The resulting  gap equations  for the scalar fields in the fundamental representation
can be written as:
\begin{equation} \label{gap-eq-Gff}  
    M_f^*=m_f -  2 G^s_{ff}\expval{\bar{\psi}_f\psi_f},
\end{equation}
where the relation of $G_{ij}$ with $G_{ff}$ and of $S_i$ with $S_f$
 are presented in  the Appendix (\ref{sec:GijGfg}).
The resulting BSE for the neutral and charged pion and kaon can be written as:
\begin{eqnarray} \label{pi-00}
\pi^0 : && 1 - \frac{G_{33}}{2} \left[ \Pi_{ps}^{uu}
 (P^2=M_{\pi^0}^2) + \Pi_{ps}^{dd} (P^2=M_{\pi^0}^2)
\right] = 0 ,
\\
K^0 :  && 
1 - G_{66}  \Pi_{ps}^{ds} (P^2=M_{K^0}^2) 
 = 0 .
\end{eqnarray}
To provide a more strict comparison with lattice calculations, below 
we also present the pion mass calculated separately with $\bar{u}u$ structure
or $\bar{d}d$ structure.
In these cases, the coupling constant $G_{33}$ was also redefined accordingly, as 
obtained in the Appendix (\ref{sec:GijGfg}) in Eq. (\ref{G33}).
It yields:
\begin{eqnarray} \label{uu-dd}
\pi^{\bar{u}u} : && 1 - \frac{G_{uu}}{2}  \Pi_{ps}^{uu}
 (P^2=M_{\pi^0}^2)  = 0 ,
\nonumber
\\
\pi^{\bar{d}d} : && 1 -  \frac{G_{dd}}{2} \Pi_{ps}^{dd}
 (P^2=M_{\pi^0}^2)  = 0 .
\end{eqnarray}

\subsection{ Mixing angles }

 The mixing type interactions $G_{08} (B)$
give rise to the eta-eta' mesons mixings.
This mixing emerges already by considering the contribution of non degenerate quark masses
$M_u \neq M_s$ \cite{PRD-2021,JPG-2022}
and we 
 we'll present exclusively the effect of the magnetic field on the mixing angle.

For this the
 auxiliary fields must be introduced in such a way to account for the mixing interactions.
This will be done by means of functional delta functions 
in  the generating functional \cite{alkofer-etal,osipov-etal}
such that in the limit of zero mixing  previous results are reproduced.
This functional delta function can be written as:
\begin{eqnarray} \label{deltafuncP}
1 \;  = \; \int   D [ P_i ] \; \delta \left( P_i - G_{ik}  j_{ps}^k \right),
\end{eqnarray}
where $j_{ps}^k = \bar{\psi} \lambda^k i \gamma_5 \psi$,
 $i,k=0,3,8$ provides   components that produce 
the mesons $\eta, \eta'$ and $\pi_0$.
Eventual non factorizations \cite{PRD-2015} 
may be expected to be  small.
Consider the following pseudoscalar auxiliary fields quadratic terms:
\begin{eqnarray} \label{P0P8-mix}
{\cal L}_{mix} &=&
- \frac{M_{88}^2(B)}{2} P_8^2  
- \frac{M_{00}^2(B)}{2} P_0^2 + 
2  G_{08} (B) \bar{G}_{08}  P_0 P_8 
+ {\cal O} (P_3, P_3^2) ...
\end{eqnarray}
where $M_{ii}^2$ include the contributions from 
$G_{i=j}$ derived above,
and 
\begin{eqnarray} \label{mix-08}
\bar{G}_{08} (B) = \frac{2
 }{  G_{00} (B)
\left( G_{88} (B) - \frac{ {G_{08} (B)}^2}{G_{00} (B) } \right)
}.
\end{eqnarray}
 The flavor dependent coupling constants $G_{ij} \propto N_c$,
 as $N_c\to \infty$,   $\eta$ and $\eta'$ become degenerate
\cite{eta-etap}.

A  change of basis state can be done   to the  mass eigenstates $\eta, \eta'$
by starting from the singlet  flavor states  basis with $|\bar{q} q>$ (q=u,d,s),
$P_3,P_8,P_0$.
By neglecting the neutral pion mixings, 
according to the 
convention from   \cite{PDG},
it can be written:
\begin{eqnarray} \label{mix-rot}
| {\eta} > &=& 
\cos \theta_{ps} | P_8 > - \sin \theta_{ps} | P_0 >,
\nonumber
\\
| {\eta}' > &=& 
\sin \theta_{ps} | P_8 > + \cos \theta_{ps} | P_0 >.
\end{eqnarray}
  To describe completely  both masses, $\eta, \eta'$
 one   needs two parameters/angles
 \cite{mix-latt}, however
in this work only the mass difference will be considered. 
 By rewriting  ${\cal L}_{mix}$ in this mass eigenstates basis,
 the following magnetic field induced
deviation of the $\eta-\eta'$ mixing angle is obtained:
\begin{eqnarray} \label{thetaps08}
\Delta  \theta_{ps} = \theta_{ps} (B)  - \theta_{ps} (B=0) =   \frac{1}{2} \arcsin \left( 
\frac{ 4  G_{0 8} (B) \bar{G}_{08} (B)  }{   (M_\eta^2 - M_{\eta'}^2 )  }  \right).
\end{eqnarray}

Let us consider $M_\eta (B=0) = 548 $ MeV and $M_{\eta'} (B=0) = 958 $ MeV
\cite{PDG}
Some results with NJL suggest  some difference in the magnetic field dependence 
of $M_\eta(B)$ and $M_{\eta'}(B)$ \cite{nonet-NJL-B}.
So we will present an estimation for the
mixing angle as a function of the magnetic field by assuming 
the following different behaviors
for the  $\eta-\eta'$ mass difference ($ \Delta_B$):
\begin{eqnarray} \label{deltaBMM}
D_{1} \equiv 
\Delta_B^{(1)} &=& \sqrt{(M_\eta^2 (B) - M_{\eta'}^2 (B))} \simeq 786 MeV \hspace{1.cm} \mbox{  is a constant},
\\
D_{2} \equiv 
\Delta_B^{(2)} &=& \sqrt{(M_\eta^2 (B) - M_{\eta'}^2 (B))} \times ( 1 + \frac{eB}{b_0} ),
\hspace{1.cm}\mbox{for} \;\;\; b_0=2 GeV^{2},
\nonumber
\\
D_{3} \equiv 
\Delta_B^{(3)} &=& \sqrt{(M_\eta^2 (B) - M_{\eta'}^2 (B))} \times ( 1 - \frac{eB}{b_0} ),
\hspace{1.cm} \mbox{ for} \;\;\;  b_0=2 GeV^{2}.
\nonumber
\end{eqnarray}

\section{ Numerical results}
\label{sec:numerics}

The result of the fitting procedure to fix the parameters of the model
with the coupling constant $G_0$ is presented 
{    in Table (\ref{tab:parameters}), where } these were the parameters found to reproduce
 $m_\pi =135.0\,\rm MeV$, $m_K=498.0\,\rm MeV$, $f_\pi=93.0\,\rm MeV$
 and $f_K=111.0\,\rm MeV$ at $B=0$.
It is interesting to note that the current quark masses fixed by the fitting procedure 
are, as usually, somewhat different than the measured values in Particle Data Group  (PDG) Tables
 \cite{PDG}.
In the NJL the Lagrangian quark masses are free parameters
and therefore they are free to be varied. However, 
it turns out that the needed values for these parameters are close to the measured values in
PDG
and this can be seen as a feature of the model.
The difference with respect to the values of PDG 
 can be attributed to, at least, two issues that might be connected.
First, the current quark masses in \cite{PDG} are fixed with respect to an energy scale of the 
Standard Model and a different energy scale may be more suitable
 for the dynamics of hadrons within the NJL.
Second, the mass generation mechanism in the NJL model involves the solution of 
(transcendental) 
gap equations for which the current quark masses contribute non-linearly, and in this
process, it might be needed a larger current quark mass that somehow is modified 
by the presence of the quark condensate.

\begin{table}[!ht]
\centering
\begin{tabular}{|c|c|}
\hline
Parameters & Set     
 \\ \hline
$\Lambda$  & $914.6\,\rm MeV$     
 \\ \hline
$G_0$        & $9.76\,\textrm{GeV}^{-2}$  
 \\ \hline
$m_{ud}$   & $6.0\,\rm MeV$     
   \\ \hline
$m_s$      & $165.7\,\rm MeV$      
       \\ \hline
\end{tabular}
\caption{
 Set of Parameters  was fixed to describe correctly neutral pion and kaon masses and 
decay constants in the vacuum ($B=0$). 
}
\label{tab:parameters}
\end{table}

The effect of the derived magnetic  field dependence of the NJL coupling constant, Eqs. (\ref{Gs-B}) and 
(\ref{Gps-B}),
 will be compared to the effect of  parameterizations considered in the literature.
For instance,
the following two shapes will be considered below \cite{avancini-etal,ferreira-etal}
\begin{eqnarray} \label{GB1}
G_1 (eB) &=& \alpha + \beta e^{ - \gamma (eB)^2},
\\
\label{GB2}
G_2 (eB) &=&
G_0 \left( \frac{ 1 +  a \xi^2  + b \xi^3}{ 1 + c \xi^2 + d \xi^4} \right)
,
\end{eqnarray}
where: 
 $\alpha = 6.70$ GeV$^{-2}$ and $\beta = 3.06$ GeV$^{-2}$ 
such that 
$G_2(0) = G_0$, and
 $\gamma = 1.31$ GeV$^{-2}$ - being 
that $\beta$ and $\gamma$ have the same values as used in Ref. 
\cite{avancini-etal}.
$G_0$ is normalized by the value presented in the Table (\ref{tab:parameters}),
$\Lambda_{QCD} = 300$ MeV, 
$\xi = (e B)/\Lambda_{QCD}^2$,and 
$a=0.01088$, $b = - 1.0133 \times 10^{-4}$, $c = 0.02228$
and $d=1.84558\times 10^{-4}$
\cite{ferreira-etal}.

\begin{figure}[ht!]
\centering
\includegraphics[width=130mm]{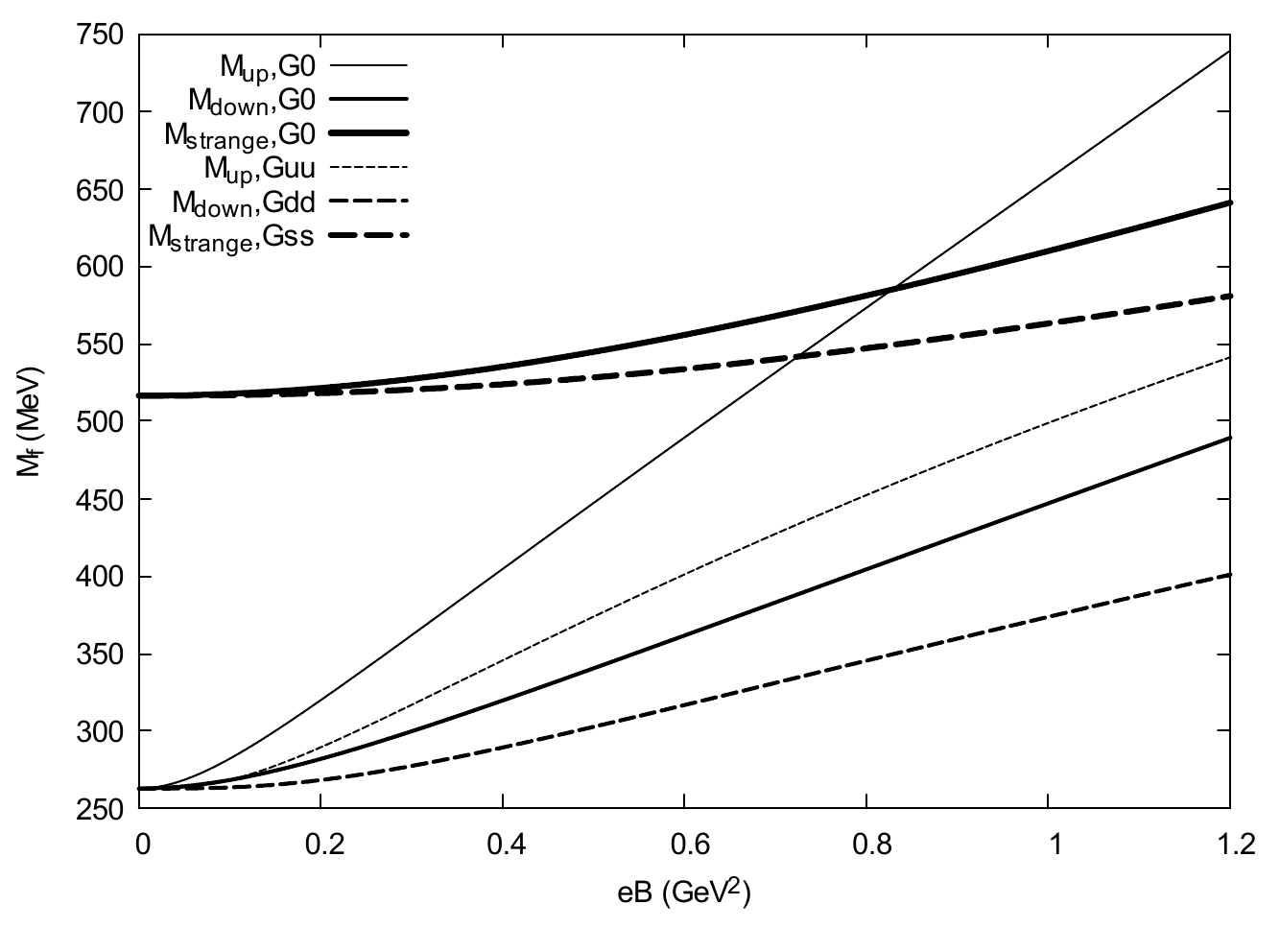}
 \caption{ \label{fig:gap1gap2}
\small
Effective quark masses as solutions of the gap equations (\ref{Gap-equation-magnetic}) 
with $G_0$ (solid lines) 
and (\ref{gap-eq-Gff}) with $G_{ff}$ (dashed lines).
Thicker (thinner) lines for $M_s^*$ ($M_u^*$) and intermediary thickness for 
$M_d^*$. 
}
\end{figure}

The quark effective masses, as solutions of the first gap equations 
for  $G_0$, Eq. (\ref{Gap-equation-magnetic}), 
and for $G_{ff}^s(B)$, the  second gap equation
 Eq. (\ref{gap-eq-Gff}), are presented in Fig. (\ref{fig:gap1gap2}).
The corrected gap equations, with $G_{ff}^s$, are solved self consistently with 
the coupling constants $G_{ij}^s(B)$.
The magnetic field corrections for the  scalar
coupling constants, that contribute in the corrected gap equations,  are shown in the following
 figure.
It is seen that the effect of the magnetic field corrections to the coupling constants  in the 
gap equations
is to reduce the effective masses.
The  deviation with respect to the solution of the gap equations
with $G_0$ is 
 progressively larger for larger magnetic fields.
The largest deviation in the effective mass is 
obtained for the up quark effective mass
and independent of  corresponding sign of the quark electric charge.

\FloatBarrier

\begin{figure}[ht!]
\centering
\includegraphics[width=130mm]{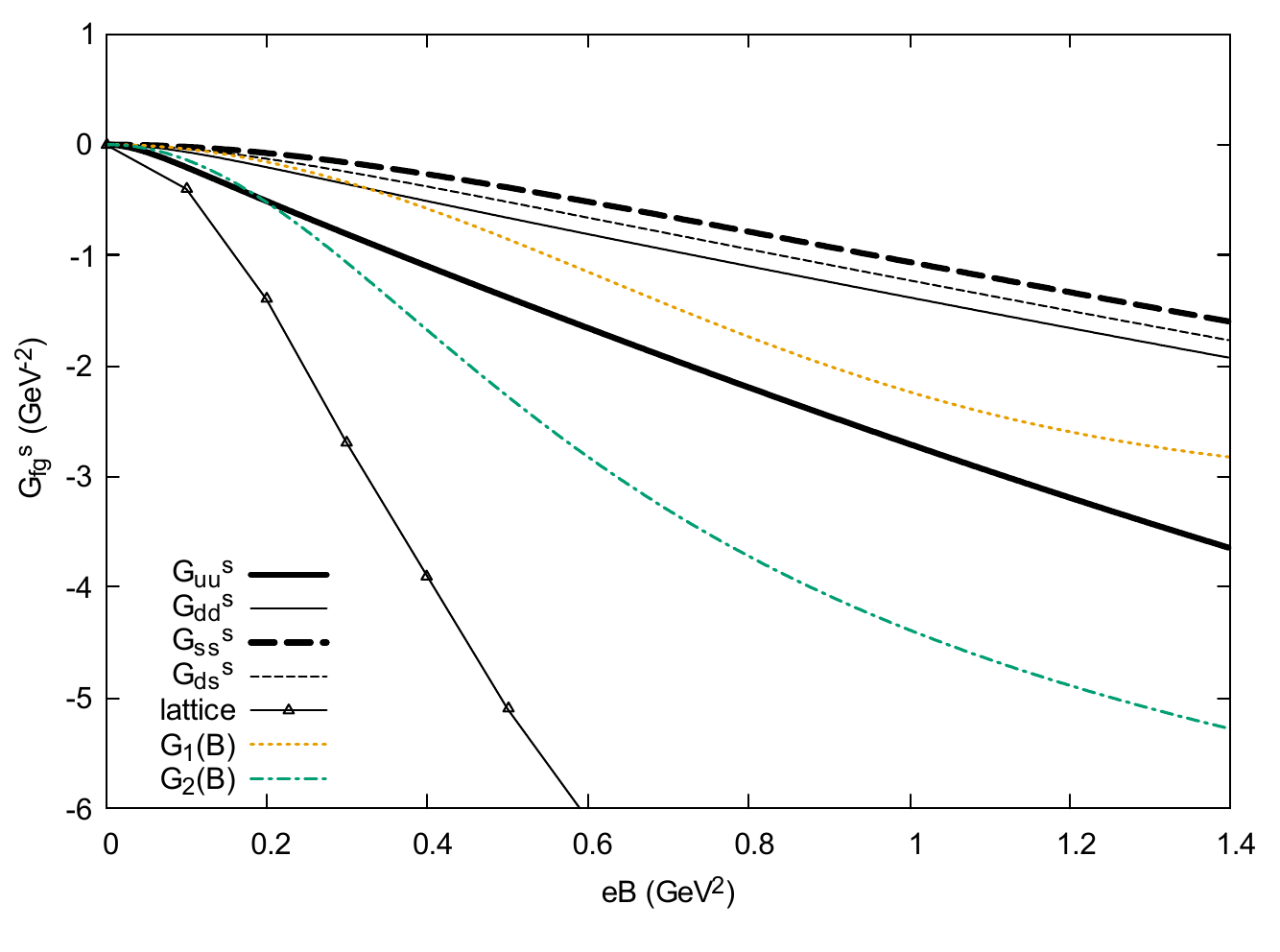}
 \caption{ \label{fig:Gs}
\small
Magnetic field correction to the scalar NJL coupling constant, $G^s_{ff}(eB)$,
  as functions of the magnetic field. 
The parameterizations of Eqs. (\ref{GB1}) and (\ref{GB2})  from Refs.
 \cite{avancini-etal,ferreira-etal} - respectively in dotted (yellow) and dot-dashed (green) lines -
and extrapolation $G(B)$ from lattice QCD of Ref. \cite{endrodi+marko} in continuous line with triangles
are also shown.
}
\end{figure}

\begin{figure}[ht!]
\centering
\includegraphics[width=130mm]{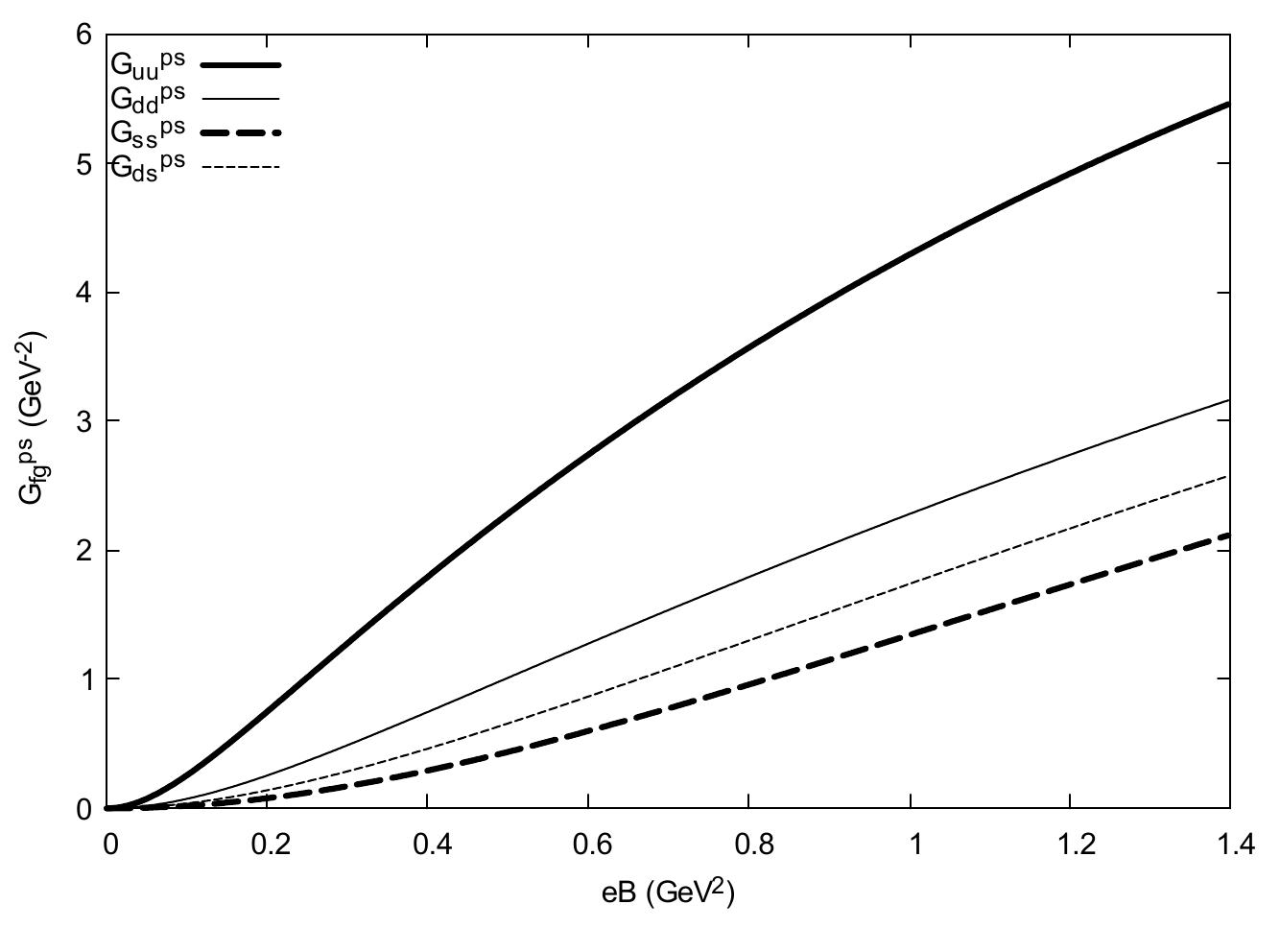}
 \caption{ \label{fig:Gps}
\small
Magnetic field correction to the pseudoscalar NJL coupling constant, $G^{ps}_{ff}(eB)$,
  as functions of the magnetic field. 
}
\end{figure}

The resulting magnetic field   dependencies of some of the
 scalar and pseudoscalar  coupling constants, 
$G_{uu}^{s,ps}(B)$, $G_{dd}^{s,ps}(B)$, $G_{ss}^{s,ps}(B)$ and  $G_{ds}^{s,ps}(B)$,
are shown in Figs. (\ref{fig:Gs}) and (\ref{fig:Gps})
 for the set of parameters above.
The pseudoscalar magnetic field corrections are positive
and they increase with the magnetic field whereas the
corrections to the  scalar coupling constant are negative and decrease with the magnetic field.
Together with the scalar coupling constants in Fig. (\ref{fig:Gs})
it is also presented:
the parameterizations
of Eqs. (\ref{GB1}) and (\ref{GB2})  from Refs.
 \cite{avancini-etal,ferreira-etal} - respectively in dotted (yellow) and dot-dashed (green) lines -
and
a set of points  for a particular definition
from lattice QCD to make contact with the NJL model  from Ref. \cite{endrodi+marko}.

\FloatBarrier

\FloatBarrier

\begin{figure}[ht!]
\centering
\includegraphics[width=130mm]{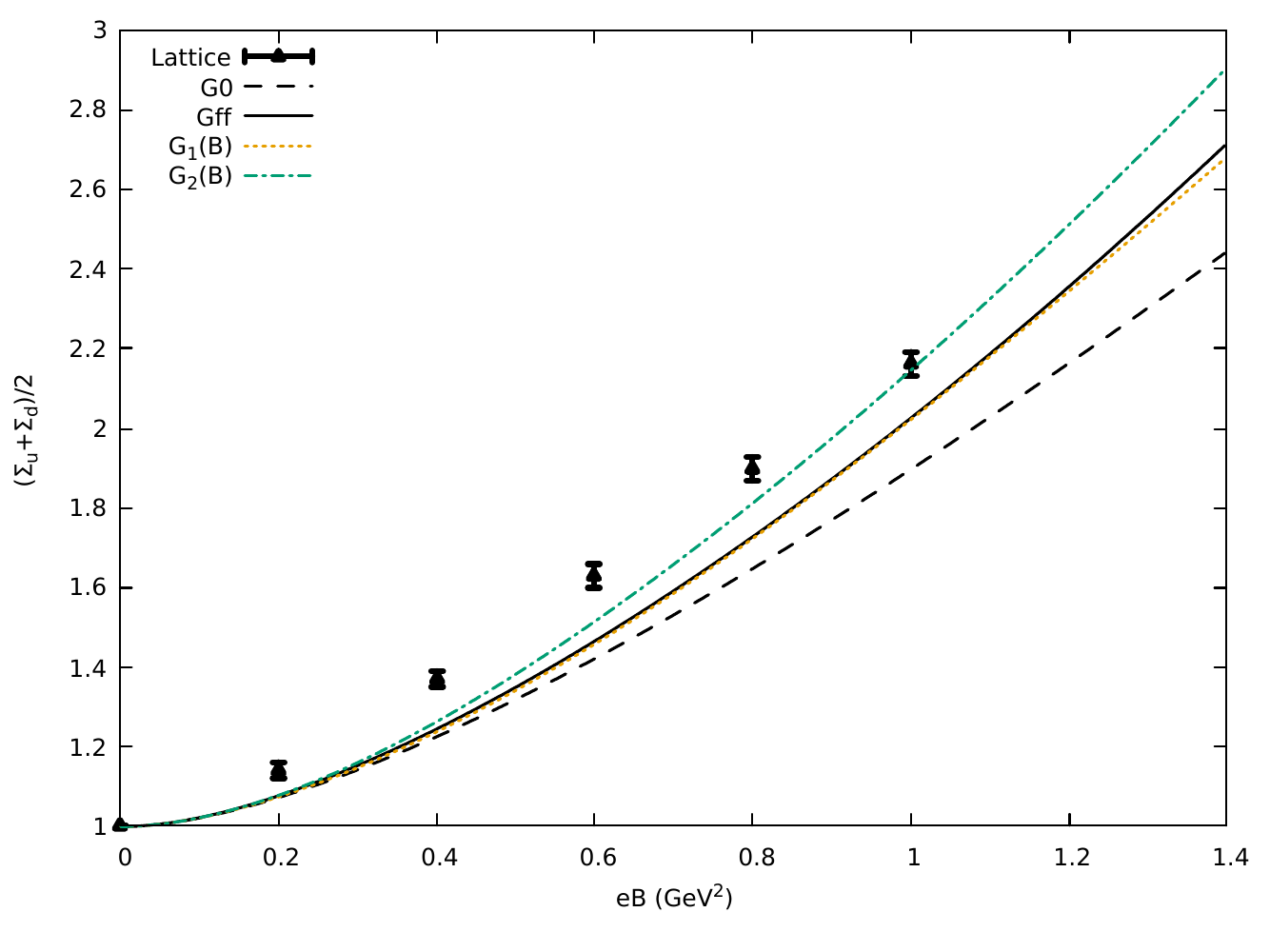}
 \caption{ \label{fig:average}
\small
Magnetic field dependent part of the averaged up and down quark condensates,
$(\Sigma_u + \Sigma_d) (B)/2$  for the set  of parameters shown above
 by using $G_0$ and $G_{ff}$,
Eqs. (\ref{Gap-equation-magnetic})  and (\ref{gap-eq-Gff}),
and also points from lattice QCD from Ref. \cite{chiral-cond}.
}
\end{figure}

The resulting up and down quark-antiquark chiral scalar condensates will be
 compared 
to results from lattice QCD from Ref. 
\cite{chiral-cond} by means of 
their average and their difference.   
To do this comparison, we define the following quantities:
\begin{equation} \label{Sigma_f(B)}
    \Sigma_f(B)=\frac{2m_{ud}}{m_\pi^2f_\pi^2}
\abs{\expval{\Bar{\psi}_f\psi_f}_B-\expval{\Bar{\psi}_f\psi_f}_{B=0}}+1,
\,\,\,\,\,\,\,\,\,\,f=u,d,s
\end{equation}
where $m_\pi$ and $f_\pi$ are the zero magnetic field pion mass and decay constant, 
respectively, here taken as $m_\pi=135\,\rm MeV$ and $f_\pi=86\,\rm MeV$.
In figure (\ref{fig:average})
the magnetic field dependent part  of the  average of the up and down quark condensates
$(\Sigma_u + \Sigma_d) (B)/2$, without the vacuum contribution, is shown
 as a  function of the magnetic field.
The curves present a comparison for the results obtained with
solutions of the two gap equations, namely for the coupling constant $G_0$ and 
for coupling constants $G_{ff}^s$,
respectively 
eq. (\ref{Gap-equation-magnetic}) 
and 
 eq. (\ref{gap-eq-Gff}).
Two different lattice calculations \cite{d-elia,htding-etal-latt}
 for these chiral condensates
present the the same behavior
with a nearly linear behavior with the magnetic field for stronger magnetic fields \cite{d-elia}.
Points obtained from lattice calculation from Ref.  \cite{chiral-cond}.
Besides that, estimations with two different parameterizations for the magnetic field
dependence of the NJL coupling constant, Eqs. (\ref{GB1}) and (\ref{GB2}), 
are presented.
The curve for the magnetic field dependent coupling constant from polarization $G_{ff}(B)$
is basically the 
same as the one from $G_1(B)$.

\FloatBarrier

\begin{figure}[ht!]
\centering
\includegraphics[width=130mm]{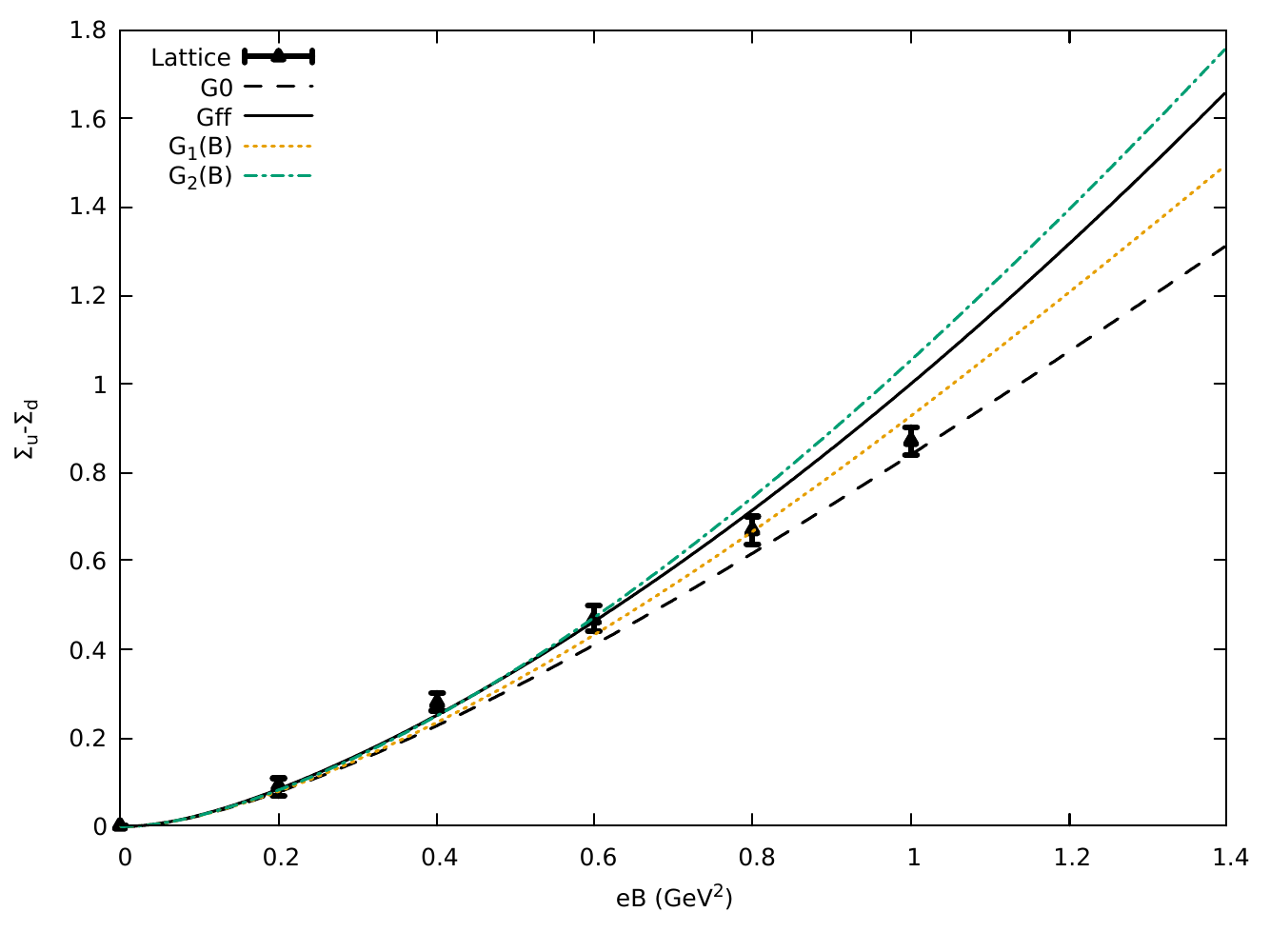}
 \caption{ \label{fig:difference}
\small
Magnetic field dependent part of the difference between the up and down quark condensates,
$(\Sigma_u - \Sigma_d) (B)$ 
  for the set of parameters shown above, by using $G_0$ and $G_{ff}$,
Eqs. (\ref{Gap-equation-magnetic})  and (\ref{gap-eq-Gff}),
and also points from lattice QCD from Ref. \cite{chiral-cond}.
}
\end{figure}

In figure (\ref{fig:difference})
the magnetic field dependent part  of the difference between the up and down quark condensates
$(\Sigma_u - \Sigma_d) (B)$, without the vacuum contribution, is shown
 as a  function of the magnetic field for the same cases presented for 
the previous figure including the lattice results from Ref. \cite{chiral-cond}.
that
It is interesting, there is an  improvement of the 
difference between the up and down quark condensates
due  to the use of $G_{ff}^s (B)$ 
with respect to the use of $G_0$ for the regime of weak magnetic field although
overall there is no systematic behavior.
Points obtained from lattice calculation from Ref.  \cite{chiral-cond}
are also shown, being  in agreement other estimations 
\cite{d-elia}.
Also, estimations with two different parameterizations for the magnetic field
dependence of the NJL coupling constant, Eqs. (\ref{GB1}) and (\ref{GB2}), 
are presented.
The resulting curve  for the coupling constant $G_{ff}(B)$ is between the curves
for $G_1(B)$ and $G_2(B)$.

\FloatBarrier

\begin{figure}[ht!]
\centering
\includegraphics[width=130mm]{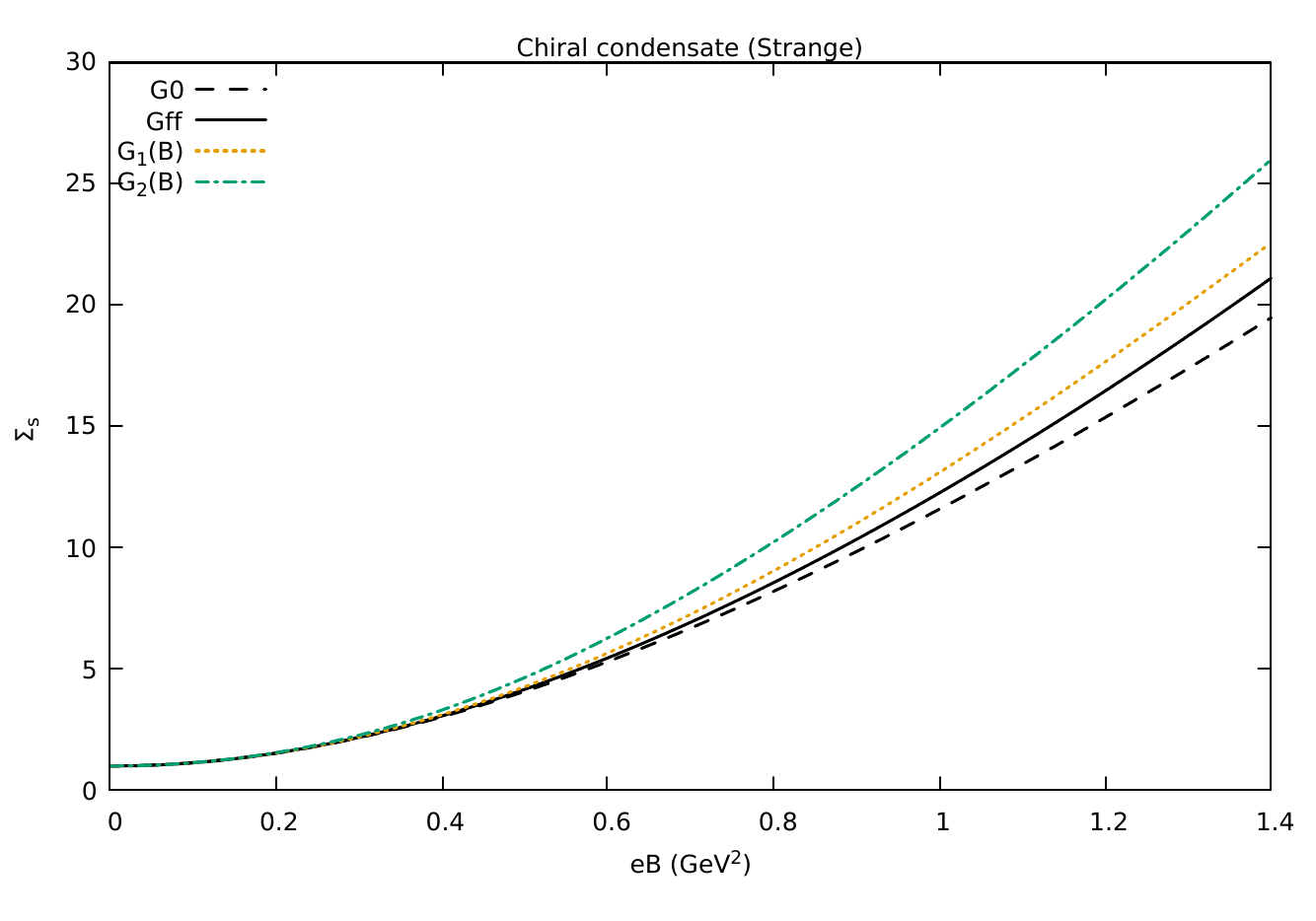}
 \caption{ \label{fig:sigma-ss}
\small
Magnetic field dependent part of the strange  quark condensate,
$\Sigma_s   (B)$  
for the solutions of the gap equation with  $G_0$ (dashed), Eq. (\ref{Gap-equation-magnetic}),
 and $G_{ss}$ (solid), Eq. (\ref{gap-eq-Gff}).
Results by considering parameterizations (\ref{GB1}) and (\ref{GB2}), respectively
dotted (yellow) and dot-dashed (green), are also shown.
}
\end{figure}

In Fig.
 (\ref{fig:sigma-ss})
 the magnetic field dependence  of the
strange  quark-antiquark condensate by means of the quantities (\ref{Sigma_f(B)})
is exhibited as functions of the magnetic field.
The case in which gap equation is solved with $G_0$ (dashed lines) and also 
the case in which the magnetic field dependent coupling constant  is used,
$G_{ff}^s(B)$ (solid lines), are shown.
For the sake of comparison, results for the two parameterizations
 (\ref{GB1}) and (\ref{GB2}), respectively
dotted (yellow) and dot-dashed (green), are also exhibited.
 The magnetic field dependent coupling constant increases the 
quark condensates, mostly because the effective masses are slightly reduced
as shown in the previous figures.
The parameterization (\ref{GB2}) yields stronger enhancement due to the magnetic field
and parameterization (\ref{GB1}) yields results nearly compatible
with the calculation with the magnetic field dependent coupling constants $G_s^{fg}$.

\FloatBarrier

\begin{figure}[ht!]
\centering
\includegraphics[width=120mm]{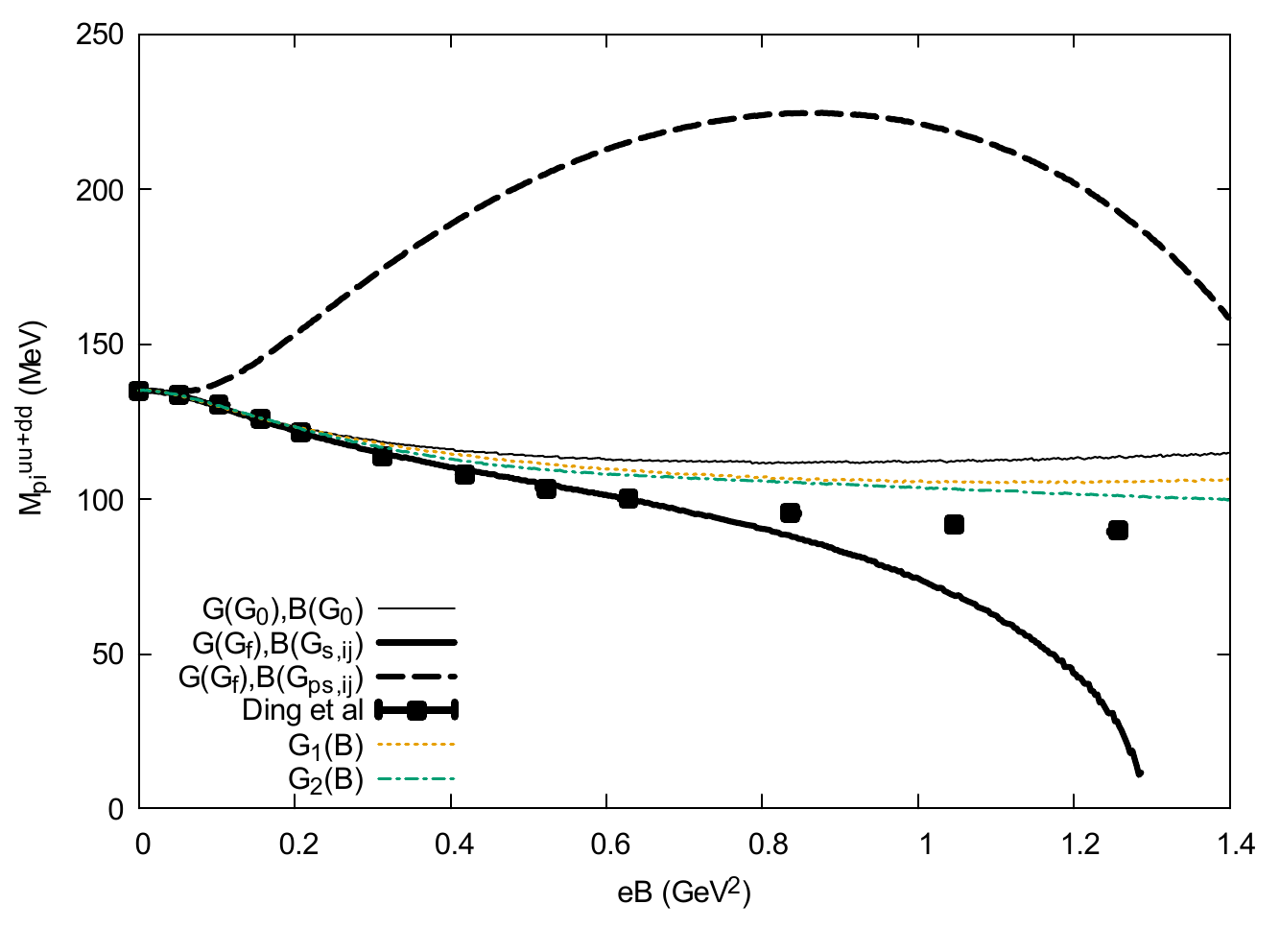}
 \caption{ \label{fig:Mpi=B=all}
\small
Neutral pion   masses (complete state) 
for the different cases discussed in the text
compared with lattice results 
from Ref. \cite{htding-etal-latt}.
}
\end{figure}

\begin{figure}[ht!]
\centering
\includegraphics[width=120mm]{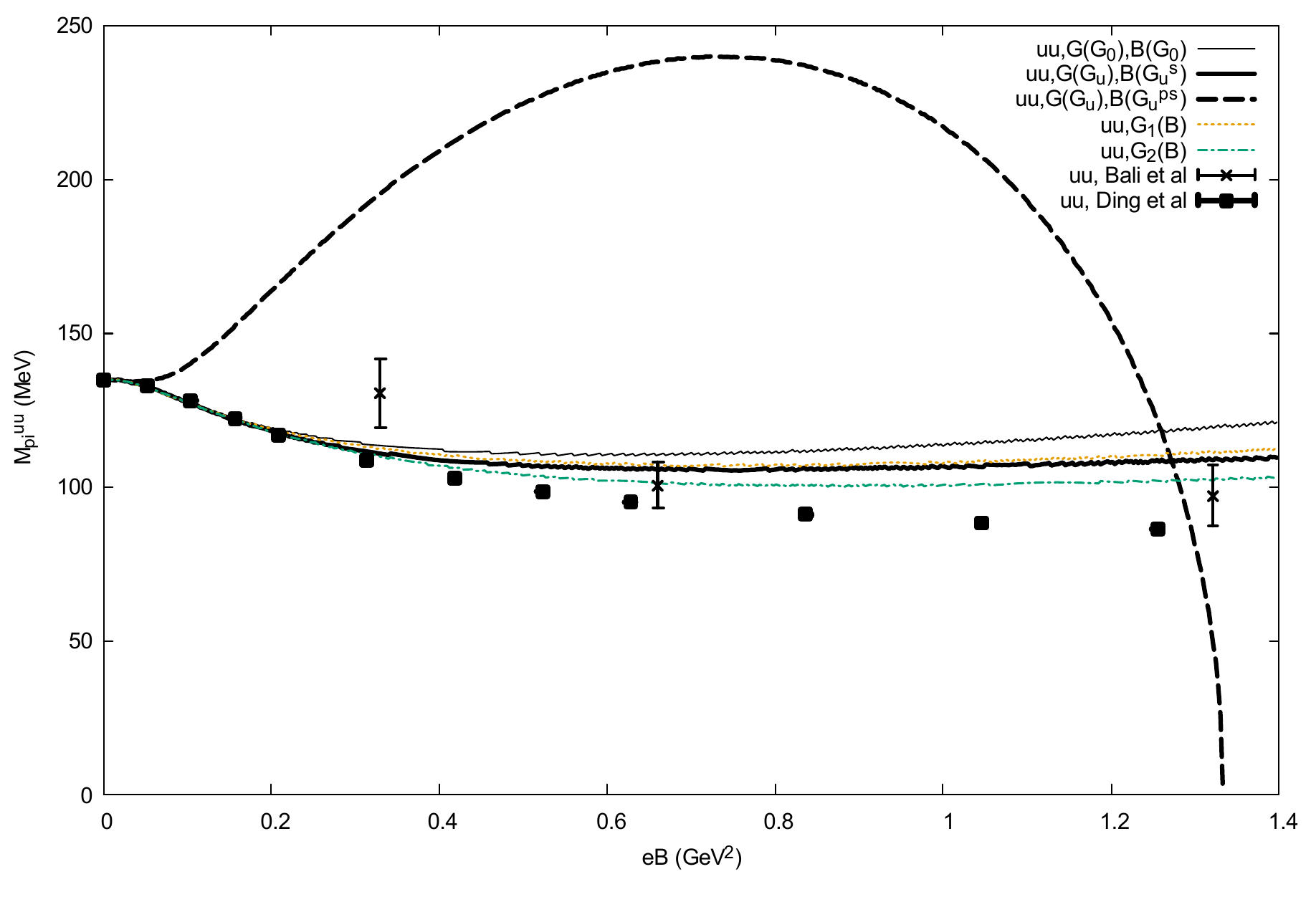}
 \caption{ \label{fig:Mpi=B=uu}
\small
Neutral pion   masses $\pi^{uu}$ for the different cases discussed in the text
compared with lattice results 
from Refs.  \cite{bali-etal-2018} and \cite{htding-etal-latt}.
}
\end{figure}

\begin{figure}[ht!]
\centering
\includegraphics[width=120mm]{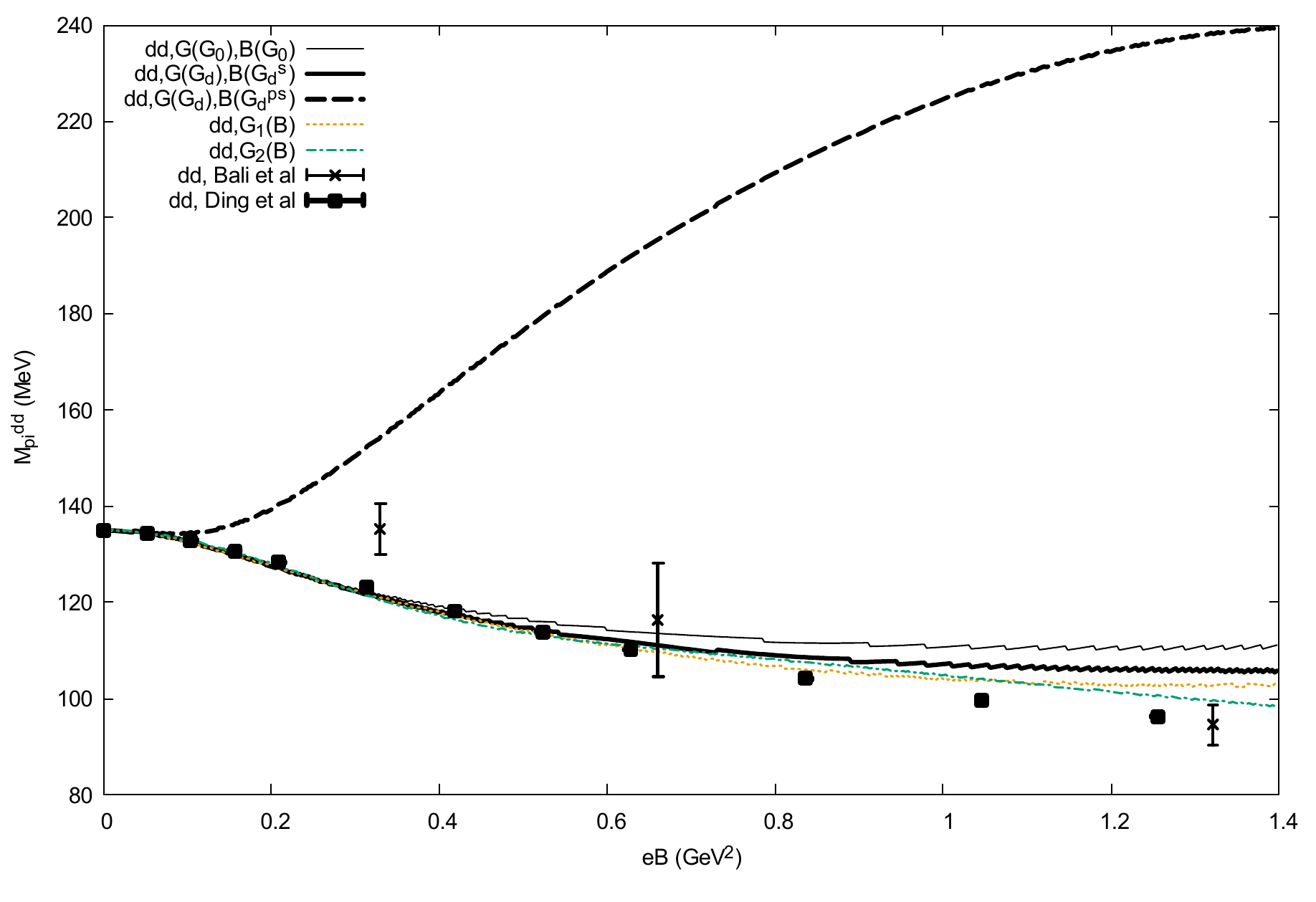}
 \caption{ \label{fig:Mpi=B=dd}
\small
Neutral pion   masses $\pi^{dd}$ for the different cases discussed in the text
compared with lattice results 
from Refs.  \cite{bali-etal-2018} and \cite{htding-etal-latt}.
}
\end{figure}

The magnetic field dependence of the three definitions of neutral pion mass are presented  in 
Figs. (\ref{fig:Mpi=B=all},\ref{fig:Mpi=B=uu}) and (\ref{fig:Mpi=B=dd}) - respectively for 
$\pi^0$ (complete pion state), $\pi^{uu}$ and $\pi^{dd}$ - see
Eqs. (\ref{pi-00}-\ref{uu-dd}).
The magnetic field behavior of the 
 pseudoscalar coupling constants does not lead to magnetic field behavior of the 
neutral pion and kaon masses
compatible with lattice estimations for strong magnetic fields. 
This can be understood by analyzing the very different behavior of 
$G^s(eB)$ and $G^{ps}(eB)$, the former is a decreasing function
of the magnetic field and the second an increasing function.
Therefore whereas $G^{s}(eB)$ yields a neutral 
pion mass that decreases with $eB$, $G^{ps}(eB)$ yields
an increasing neutral pion mass, as shown below.
To make possible a more detailed comparison of the  effect of the magnetic field
dependent coupling constants 
the pion mass was calculated in different ways.
First for the effective mass from the gap equations ($G(G_0)$) and BSE  ($B(G_0)$)
by considering $G_0$.
Secondly for the gap equations ($G$) with $G_{ff}^s(B)$ and 
BSE ($B$) with $G_{33}^s(B)$.
The same result obtained by considering $G_{ff}^{ps}(B)$ was also plotted.
The two parameterizations of Eqs. (\ref{GB1}) and (\ref{GB2}) were also employed.
Besides that, in most Lattice QCD
 the pion masses have been calculated 
 for the separated states $\pi^{\bar{u}u}$ and $\pi^{\bar{d}d}$
 as discussed for Eq. (\ref{uu-dd}).
Lattice QCD results from Refs.  \cite{bali-etal-2018} and \cite{htding-etal-latt}
 are also  shown in the figures below.
The magnetic field coupling constants reduce the 
values of the pion mass in all cases.
The pion mass decreases
also  because of the behavior of the 
quark effective masses.
It is noted the considerable role of magnetic field coupling constant for the 
complete neutral pion state mass (thick solid line) 
that reduces the pion mass  with respect to the unique coupling constant $G_0$
(thin solid line) in  Fig. (\ref{fig:Mpi=B=all}).
However due to  the non-linearity of the BSE
and to  the behavior of the quark effective masses with $eB$
the pion mass (complete state) drops too fast for magnetic fields
stronger than $eB \sim 0.65$GeV$^2$.
Eventually the pion mass go to zero by $eB \sim 1.3 GeV^2$,
 and there is no more solution for the corresponding
BSE.
For the cases of $\pi^{\bar{u}u}$ and $\pi^{\bar{d}d}$ states,
exhibited in Figs. (\ref{fig:Mpi=B=uu}) and (\ref{fig:Mpi=B=uu}) respectively,
for lower values of magnetic fields the 
magnetic field dependent scalar $G^s_{ij}(eB)$ improves the agreement 
with lattice QCD. 
However the magnetic field dependent coupling constants
$G_{ff}(B)$ are not enough to reproduce lattice QCD data for quite strong magnetic fields, 
nearly at the same point for
  the complete pion state and for the $\bar{u}u$ or $\bar{d}d$ states.
Note that the BSE for the complete neutral pion state is not consistent
with an assumption such that the complete neutral pion mass would be the 
average of the $\bar{u}u$ and $\bar{d}d$ states.
When comparing the BSE Eqs. (\ref{pi-00}) and (\ref{uu-dd}) for the complete and 
 $\bar{u}u$/$\bar{d}d$ pion states it can be seen that $G_{33}$ is an averaged of 
$G_{uu}$ and $G_{dd}$, and  also $\Pi^{complete}(P^2=M_{\pi}^2)$ is an average of the 
separated polarization tensors for u and d quarks. Since all the polarization tensors
are non-linear functions of the pion mass (in the limit of zero pion 3-momentum),
it turns out that the two averages taken to compute the complete pion mass 
in $G_{33} \Pi (M_{\pi}^2)$ varies considerably faster than the separated quantities
$G_{uu} \Pi_{uu} (M_\pi^2)$ and $G_{dd} \Pi_{dd} (M_\pi^2)$.
These behaviors lead to the unexpected faster variation of the 
complete pion mass with the magnetic field.
Of course the separated dependencies of all the three
 polarization tensors on $eB$  and on $P^2=M_{\pi}^2$
produce this unexpected behavior.
 However 
further investigation is seemingly needed to 
certify, first of all,
 different lattice calculations   provide results in agreement with each other.

\FloatBarrier

\FloatBarrier

\FloatBarrier

\begin{figure}[ht!]
\centering
\includegraphics[width=120mm]{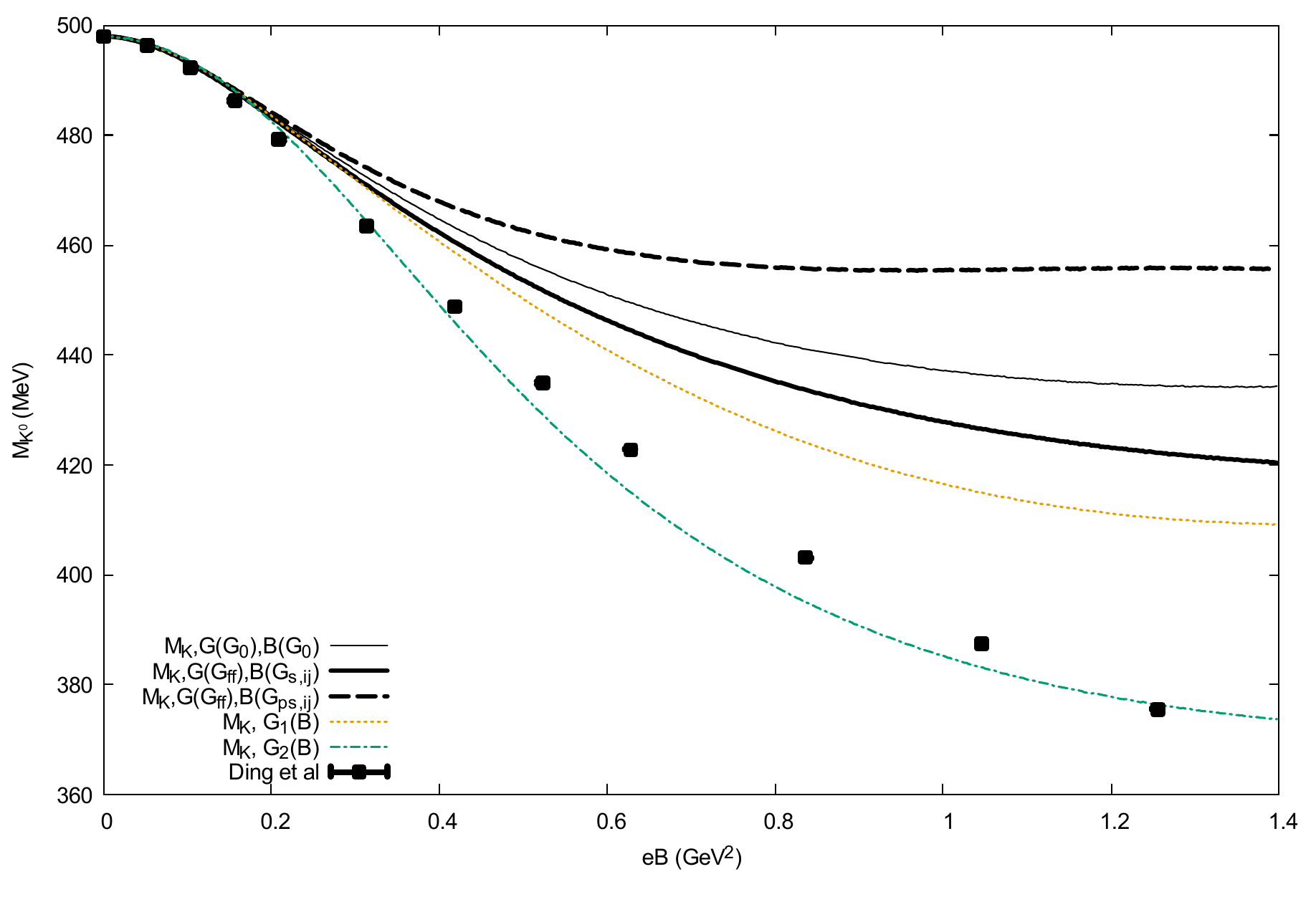}
 \caption{ \label{fig:Mka}
\small
Neutral   kaon masses for the following cases:
gap eqs.  and BSE with  $G_0$ (thin line), and 
 gap eqs. and BSE respectively with $G_{ff}^s$ and $G_{66}^s$  (thick line).
The use of $G_{66}^{ps}$ is also considered (dashed line).
By using
 $G_1(B)$ and $G_2(B)$ results are exhibited respectively
in  dotted yellow line 
and dot-dashed green line.
}
\end{figure}

In Fig. (\ref{fig:Mka}) the magnetic field dependence of the neutral kaon mass
is presented for the different  cases discussed above:
by using gap equations and BSE  with $G_0$ (thin solid line)
and gap equations with $G_{ff}^s$ and BSE with $G_{66}^s, G_{77}^s$
(thick solid line)
and also $G_{66}^{ps}, G_{77}^{ps}$ (dashed line).
The parameterizations (\ref{GB1}) and (\ref{GB2}) were also 
used, respectively dotted (yellow) and dot-dashed (green) lines.
The pseudoscalar coupling constant $G_{ij}^{ps}(eB)$ 
does not make neutral kaon mass to increase, as it happens in the
neutral pion case, although it makes results worsen when compared to results with $G_0$.
It is seen that the magnetic field  deviation due to the magnetic field 
dependent coupling constant is not enough to reproduce lattice QCD results
although it improves agreement when compared with results obtained with $G_0$.

\FloatBarrier

\begin{figure}[ht!]
\centering
\includegraphics[width=130mm]{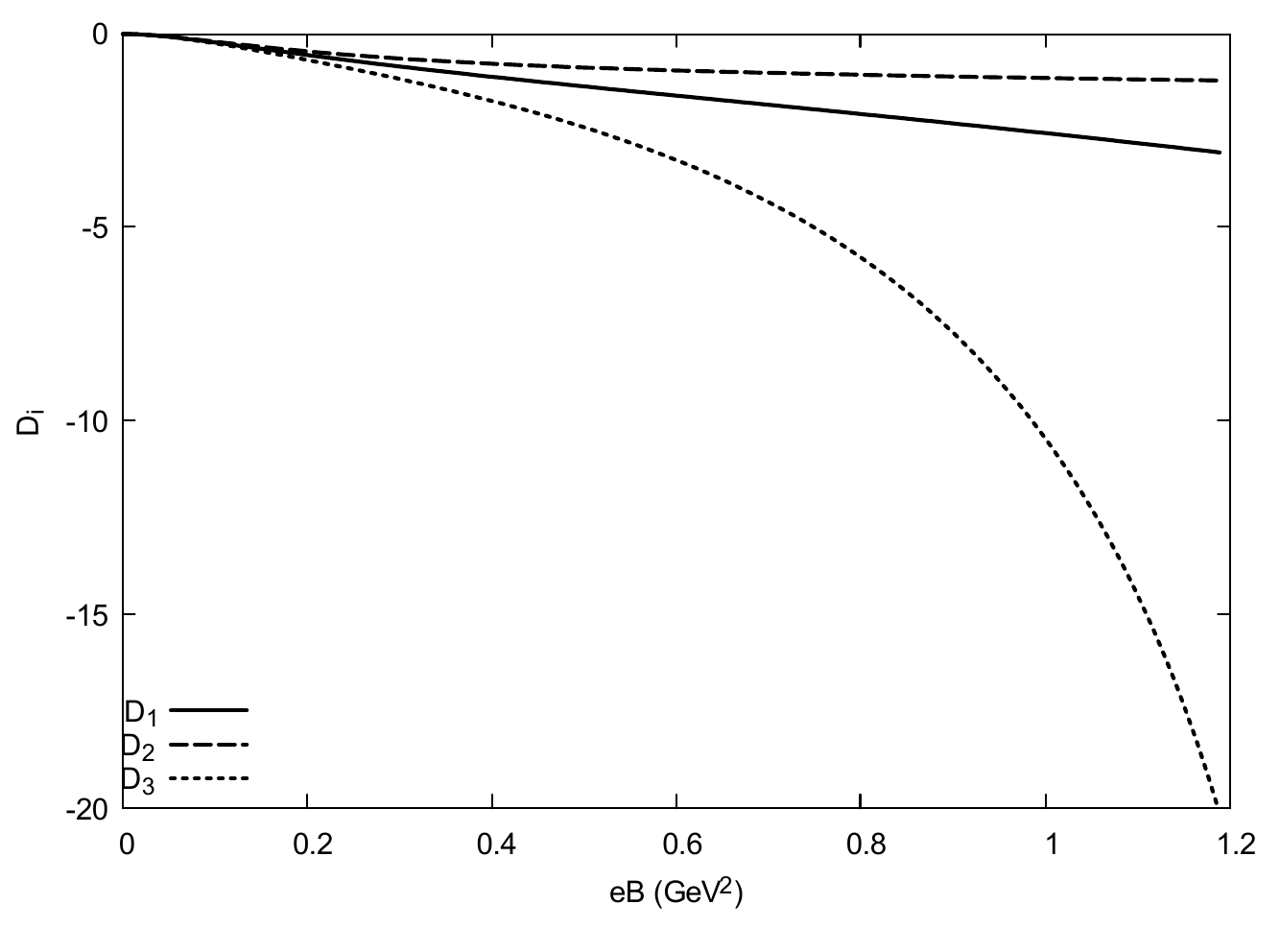}
 \caption{ \label{fig:mixingangle}
\small
Magnetic field induced deviation for the $\eta-\eta'$ mixing angle given by Eq. (\ref{thetaps08})
and the three prescriptions for the $\eta-\eta'$ mass differences of Eq. (\ref{deltaBMM}).
}
\end{figure}

In Fig. (\ref{fig:mixingangle}) the deviation of the $\eta-\eta'$ 
mixing angle due to the magnetic field, Eq. 
 (\ref{thetaps08}), 
is presented for three different ad hoc prescriptions for the behavior of the
$\eta-\eta'$ mass difference with the magnetic field shown in Eq. (\ref{deltaBMM}).
Again the coupling constants $G_{ij}^s$ were used.
The decrease of the $\eta-\eta'$ mass difference, $D_3$, contributes for a 
further increase of the modulus of the mixing angle that is favored by an increase of the 
coupling constant $G_{08}(B)$ with the magnetic field. 
The magnetic field dependencies of $G_{00}(B)$ and $G_{88}(B)$  are 
 less relevant  than $G_{08}(B)$ for the 
resulting mixing angle.
Results with the use of prescription $D_3$ are more sensitive 
to the magnetic field because $D_3$ considers a reduction of the mass different with
the magnetic field in the argument of the  arcsin
in Eq. 
 (\ref{thetaps08}).

\FloatBarrier

\section{Summary and Discussion}
\label{sec:summary}

Effects of quark polarization in a constant background magnetic field
 on the NJL-coupling constant were analyzed firstly
in the resulting gap equations,
 and therefore in the quark-antiquark chiral condensates, and 
 mass generation for  constituent quarks.
Secondly their effects were analyzed in 
the BSE for  the neutral pion and kaon masses and the $\eta-\eta'$ mixing angle.
The one loop level calculation under 
magnetic field
 breaks  chiral and  flavor symmetries inducing
different contributions for the scalar and pseudoscalar channels and 
 flavor dependency of the coupling constants.
Besides the diagonal coupling constants $G_{ii}$, mixing type interactions $G_{i\neq j}$ (for $i,j=0,3,8$)
also emerge and they  contribute to neutral  mesons mixings.
These mixing interactions have two sources:  the magnetic field
coupling to quarks and  the non degenerate quark masses, being this second
effect was also  analyzed separately in
Refs. \cite{PRD-2021}.
The resulting mixing-type interactions
are proportional to the different quark mass differences,
$\propto (M_f-M_g)$  and $(M_f^2-M_g^2)$
for $f\neq g = u,d,s$,
 and they
 were mostly  considered for an estimate
of the magnetic field correction to the $\eta-\eta'$ mixing angle.
The magnetic field dependence of the up and down quark-antiquark condensates
from the gap equations depend 
on the scalar coupling constants, $G_0 + G_{ii}^s$, 
and these results can be said to be slightly  improved with respect to results
available  from lattice QCD  calculations  although the averaged value may be
well reproduced.
It indicates, however, that further flavor or magnetic field-dependencies of parameters
may be needed mainly
to reproduce correctly the lattice results for the
difference of the up and down quark condensates.
The strange quark-antiquark condensate also receives corrections.

Although the {\it corrected}  scalar coupling constants 
 have a magnetic field dependence
with nearly  the same behavior of the coupling constant behavior needed to reproduce
lattice QCD results,
the {\it corrected} pseudoscalar coupling constants  in this one loop fermion calculation, 
$G_0 + G_{ii}^{ps}$,
has the opposite magnetic field dependence and they do not lead to results with the behavior 
found in lattice QCD results.
Therefore the pseudoscalar coupling constants were not employed 
 extensively for calculating observables.
This  suggests that  
there may  have a  further different mechanism
 in the pseudoscalar channel
that could generate a strongly decreasing behavior for $G_{ij}^{ps} (B)$ that 
should compensate the behavior obtained from polarization process.
Therefore, by simply 
adopting the scalar coupling constant 
to compute the neutral pion   bound states,
 results receive corrections that somewhat
improve the agreement with 
data from lattice QCD.
This comparison  presents some subtleties because lattice QCD  
 calculations
have few points for finite magnetic field and they
 provided neutral pion mass  mostly for separated 
$\bar{u}u$ or $\bar{d}d$ structures.
Therefore, to make possible a more detailed comparison
among different calculations, we presented calculations for the 
complete neutral pion state mass 
and for the  $\bar{u}u$ or $\bar{d}d$ states.
Neutral pion mass as calculated for $G_0$
 and  for the separated states   $\bar{u}u$ or $\bar{d}d$
present a similar behavior: for lower magnetic fields there is a decrease of the 
masses and
NJL-predictions yield,   for $eB \ge 0.5-0.9$GeV$^{2}$, an increase of masses.
A different behavior is obtained for the complete neutral pion structure
for $G_{ij}^s(B)$ with a continuous decrease of its mass
until there is no more solution for the neutral pion BSE 
around $eB \sim 1.3$GeV$^2$.
Note that  the complete neutral pion mass is not an average of the 
masses of states $\bar{u}u$ and $\bar{d}d$ because of the non linearity of the BSE 
but also due to the different up and down quark effective masses.
It is interesting to emphasize that whereas the current NJL predictions for the
 up and down quark condensates
are rea improved with respect to the standard NJL 
the results for the neutral mesons masses need further physical input in their BSE.

The neutral kaon mass calculated either  with $G_0$ or with $G_{ij}^s$ provide
decreasing values with $eB$ although the magnetic field dependent coupling constants
provide stronger decrease.
By $eB_0 \sim 1.0$GeV$^2$, the difference between the two estimates
is of the order of
$M_{K^0}(G_0) - M_{K^0}(G_{ij}(B)) \sim 10$MeV,
and larger for stronger magnetic fields.
Finally estimates for the magnetic field dependence of 
the   $\eta-\eta'$ mixing angle were provided by 
considering the mixing type interaction $G_{08}(B)$ according to 
Refs. \cite{PRD-2021,JPG-2022}.
As shown in the Appendix (\ref{sec:GijGfg})
$G_{08}(B) \sim G_{uu} + G_{dd} - 2 G_{ss}$ that is proportional to
the up/down -strange quark effective mass non-degeneracy.
For the $\eta-\eta'$ 
mixing angle, different behaviors of the magnetic field dependence of the 
 mass difference $M_{\eta'} - M_{\eta}$(B) were considered.

These results suggest that
the present magnetic field corrections for the NJL coupling constant
from quark-polarization might be enough to describe results for the neutral
pion  mass from lattice QCD  for not strong magnetic fields, i.e. 
$eB \lesssim 0.2, 0.4$ or $0.6$ GeV$^{2}$, depending on the definition of the pion struture
according to Figs.  (\ref{fig:Mpi=B=all},\ref{fig:Mpi=B=uu}) and (\ref{fig:Mpi=B=dd}) .
Neutral kaon masses are also well reproduced for still weaker magnetic fields.
The higher order  polarization corrections should not provide large contributions
because they are suppressed by $1/{M^*}^n$ ($n \geq 2$).
Therefore, further magnetic field dependencies might be needed for 
realistic predictions of the NJL model.
Further comparisons of NJL predictions with first principles lattice QCD results
will make possible to understand better, and eventually to improve, the predictive power of the model
under finite magnetic fields.
For that it is also important to provide further 
  lattice calculations.
Nevertheless, with calculations presented in this work, it is possible to  identify 
how the NJL-degrees of freedom 
-exclusively  - come into play for the corresponding hadron observables
under finite magnetic fields.
This procedure should help to disentangle somewhat both the understanding
of hadron dynamics in terms of the fundamental degrees of freedom and 
in terms of hadron effective (and observable) degrees of freedom by trying to relate both 
levels of the description.
Maybe, 
this type  of comparisons
 also might eventually help  to conclude further which "sector" of QCD dynamics is at work
for each observable
under these external conditions.

\section*{Acknowledgments}

The authors  thank short conversations with 
I. Shovkovy and G. Mark\'o,
and G. Endrodi and H.T. Ding for 
 sending tables with results obtained in Lattice QCD of their groups.
F.L.B. is member of
INCT-FNA,  Proc. 464898/2014-5.
T.H.M. thanks financial support from CAPES.
F.L.B. thanks partial support from 
CNPq-312072/2018-0, 
CNPq-421480/2018-1 and CNPq-312750/2021-8.

\numberwithin{equation}{section}

\appendix

\section{ Quark propagator in a constant magnetic field}
\label{sec:q-prop}

By considering the  proper time representation for 
the quark propagator  with the 
minimal coupling  to the photon field is given by:
\begin{equation}
    S_0(x,y)=\Phi(x,y)\mathcal{S}_0(x-y),
\end{equation}
where
\begin{equation} \label{Phi}
    \Phi(x,y)\equiv \exp\left\{ iq\int_y^x d\xi^\mu\,\left[ A_\mu (\xi)+\frac{1}{2}F_{\mu\nu}\left( \xi-y\right) ^\nu \right] \right\} ,
\end{equation}
is the Schwinger phase factor, which is explicitly gauge dependent and breaks the translation invariance of the propagator, and
\begin{equation}
\begin{split}
    \mathcal{S}_0(x-y)\equiv-\left( 4\pi\right) ^{-2}\int_0^\infty \frac{ds}{s^2}\left[m + \frac{1}{2}\gamma\cdot\left[ q\mathbf{F}\coth\left( q\mathbf{F}s\right) +q\mathbf{F}\right] (\mathbf{x}-\mathbf{y}) \right] \\
    \times\exp\left\{ -im^2 s-\frac{1}{2}\textrm{tr}\ln\left[ \left( q\mathbf{F}s\right) ^{-1} \sinh\left( q\mathbf{F}s\right) \right] \right\} \\
    \times\exp\left[-\frac{i}{4}(\mathbf{x}-\mathbf{y})^Tq\mathbf{F}\coth\left( q\mathbf{F}s\right) (\mathbf{x}-\mathbf{y}) +\frac{i}{2}q\sigma_{\mu\nu}F^{\mu\nu}\,s \right] 
\end{split}
\end{equation}
is the translational invariant term. Here the quark electric charge is denoted by $q$ while $m$ stands for its mass. The photon field strength tensor is denoted by $F^{\mu\nu}$ and $\sigma^{\mu\nu}=\frac{i}{2}\comm{\gamma^\mu}{\gamma^\nu}$.

Now we consider the case in which the photon field correspond to a constant
magnetic field along the $\hat{z}$ direction, $\vec{B} = B \hat{e}_z$,
such that $F_{12} = B$. In this case, the translational invariant propagator becomes
\begin{equation} \label{Bz-finaleq}
\begin{split}
    \mathcal{S}_0(x-y)=-\left( 4\pi\right) ^{-2}\int_0^\infty \frac{ds}{s^2}\frac{\abs{qB}s}{\sin\left( \abs{qB}s\right)}\exp\left( -im^2 s+i\,\textrm{sign}(qB)\abs{qB}s\vb*{\sigma}_3\right) \\
    \times\exp\left\{-\frac{i}{4s}\left[ (x-y)_\parallel^2-\abs{qB}s\cot\left( \abs{qB}s\right) (x-y)_\perp^2\right] \right\} \\
    \times\left\{ m+\frac{1}{2s}\left[ \gamma\cdot (x-y)_\parallel -\frac{\abs{qB}s}{\sin \left( \abs{qB}s\right)}\gamma\cdot (x-y)_\perp e^{-i\,\textrm{sign}(qB)\abs{qB}s\vb*{\sigma}_3}\right] \right\} ,
\end{split}
\end{equation}
where $\textrm{sign}(x)$ is the sign function and, for two arbitrary $4-$vectors $a^\mu$ and $b^\mu$, we are denoting
\begin{align*}
    \qty(a\cdot b)_\parallel&=a^0b^0-a^3b^3, \\
    \qty(a\cdot b)_\perp&=a^1b^1+a^2b^2.
\end{align*}

The Fourier transformation of eq. (\ref{Bz-finaleq}) is found to be given by
\begin{equation}
    \begin{split}
        \mathcal{S}_0(p)=-i\int_0^\infty ds\,\exp\qty{-is\qty[m^2-p_\parallel^2+\frac{\tan(\abs{qB}s)}{\abs{qB}s}p_\perp^2]} \\
        \times\qty{\qty[1-\textrm{sign}(qB)\gamma_1\gamma_2\tan(\abs{qB}s)]\qty(m+\gamma\cdot p_\parallel)-\gamma\cdot p_\perp\qty[1+\tan^2\qty(\abs{qB}s)]}.
    \end{split}
\end{equation}

\section{ Coupling constants in different flavor basis and  Integrals} 
\label{sec:GijGfg}

\numberwithin{equation}{section}

The coupling constants of NJL interaction in the adjoint representation relates to the ones in the fundamental representation by

\begin{subequations} \label{Adjoint-to-fundamental}
\begin{equation}
    G^{00}=\frac{1}{3}\qty[G_{uu}(B)+G_{dd}(B)+G_{ss}(B)],
\end{equation}
\begin{equation} 
    G^{11}(B)=G^{22}(B)=G_{ud}(B),
\end{equation}
\begin{equation} \label{G33}
    G^{33}(B)=\frac{1}{2}\qty[G_{uu}(B)+G_{dd}(B)],
\end{equation}
\begin{equation}
    G^{44}(B)=G^{55}(B)=G_{us}(B),
\end{equation}
\begin{equation}
    G^{66}(B)=G^{77}(B)=G_{ds}(B),
\end{equation}
\begin{equation}
    G^{88}(B)=\frac{1}{6}\qty[G_{uu}(B)+G_{dd}(B)+4G_{ss}(B)],
\end{equation}
\begin{equation}
    G^{03}(B)=G^{30}(B)=\frac{1}{\sqrt{6}}\qty[G_{uu}(B)-G_{dd}(B)],
\end{equation}
\begin{equation}
    G^{08}(B)=G^{80}(B)=\frac{1}{3\sqrt{2}}\qty[G_{uu}(B)+G_{dd}(B)-2G_{ss}(B)],
\end{equation}
\begin{equation}
    G^{38}(B)=G^{83}(B)=\frac{1}{2\sqrt{3}}\qty[G_{uu}(B)-G_{dd}(B)],
\end{equation}
\end{subequations}
both for scalar and pseudoscalar interactions. All the other couplings $G^{ij}$ vanish. Here we are denoting
\begin{subequations}
\begin{equation}
    G_{fg}^{\,\rm s}(B)=g+g^2\Pi_{fg}^{\,\rm s}(B),
\end{equation}
\begin{equation}
    G_{fg}^{\,\rm ps}(B)=g+g^2\Pi_{fg}^{\,\rm ps}(B),
\end{equation}
\end{subequations}
where
\begin{subequations}
\begin{equation}
    \Pi_{fg}^{\,\rm s}(B)=2iN_c\int\frac{d^4p}{(2\pi)^4}\textrm{tr}_D\qty[S_f^B(p)S_g^B(p)],
\end{equation}
\begin{equation}
    \Pi_{fg}^{\,\rm ps}(B)=2iN_c\int\frac{d^4p}{(2\pi)^4}\textrm{tr}_D\qty[S_f^B(p)i\gamma_5S_g^B(p)i\gamma_5],
\end{equation}
\end{subequations}
with $S_f^B(p)$ representing the quark propagator in momentum space in the presence of the uniform magnetic field $B$ and $\textrm{tr}_D$ representing the trace over Dirac indices.

\end{document}